\documentclass[12pt]{article}
\usepackage{amsmath}
\usepackage{amsfonts}
\usepackage{amssymb}
\usepackage{graphicx}

\input{tcilatex}

\begin{document}

\title{{\Large \textbf{The manifest covariant soliton solutions on
noncommutative orbifold $T^{2}/Z_{6}$ and $T^{2}/Z_{3}$}}}
\author{Hui Deng, \thanks{%
Email:hdeng\_phy@yahoo.com.cn}\hspace{5mm} Bo-Yu Hou,\thanks{%
Email:byhou@nwu.edu.cn} \hspace{5mm} Kang-Jie Shi, \thanks{%
Email:kjshi@nwu.edu.cn} \and \hspace{5mm} Zhan-Ying Yang, \thanks{%
Email:zyyang@nwu.edu.cn}\hspace{5mm} Rui-Hong Yue,\thanks{%
Email:rhyue@nwu.edu.cn} \\
{\footnotesize Institute of Modern Physics, Northwest University,}\\
{\footnotesize Xi'an, 710069, P. R. China }}
\maketitle

\begin{abstract}
In this paper, we construct a closed form of projectors on the integral
noncommutative orbifold $T^2/Z_6$ in terms of elliptic functions by $GHS$
construction. After that, we give a general solution of projectors on $%
T^{2}/Z_{6}$ and $T^{2}/Z_{3}$ with minimal trace and continuous reduced
matrix $M(k,q_{0})$.The projectors constructed by us possess symmetry and
manifest covariant forms under $Z_{6}$ rotation. Since projectors correspond
to the soliton solutions of field theory on the noncommutative orbifold, we
thus present a series of corresponding manifest covariant soliton solutions.
\end{abstract}


\bigskip

\section{Introduction}

The idea that the space-time coordinates do not commute is quite old \cite%
{synder}. Indeed, noncommutative geometry has arisen in at least three
distinct but closely related contexts in string theory. Witten's open string
field theory formulates the interaction of bosonic open strings in the
language of noncommutative geometry \cite{6}. Compactification of matrix
theory on noncommutative tori was argued to correspond to supergravity with
constant background three form tensor field \cite{4}. More generally, it has
been realized that noncommutative gauge theory arises in the world-volume
theory on D-brane in the presence of a constant background $B$ field in
string theory \cite{5}. Until now people have made a lot of contribution to
the mathematics and physical application of noncommutative geometry.\cite%
{1,2,3}

Naturally One would like to know what's new that arises from the quantum
field theories on noncommutative space. The $UV/IR$ mixing caused by
noncommutativity of space-time is one of the intriguing aspects of
noncommutative field theory\cite{minwalla,toumbas}. Noncommutative field
theory provides us with a lively description about the quantum Hall effect%
\cite{gubser,poly,hellerman}. The research about the quantum Hall effect
concentrates plenty of interest\cite{susskind}-\cite{17}. As an important
object soliton solution always abstracts a lot of concern of string
theorists. Although Derrick's theorem forbids solitons in ordinary more than 
$1+1$ dimensions scalar field theory\cite{Derrick}, \ however Gopakumar,
Minwalla and Strominger pointed out that there exit soliton solutions in
noncommutative\ scalar field theory\cite{strominger}. It was soon realized
that noncommutative solitons represent D-branes in string field theory with
a background $B$ field\cite{dasgupta,harvey}, and many of Sen's conjectures 
\cite{sen1,sen2} regarding tachyon condensation in string field theory have
been beautifully confirmed using properties of noncommutative solitons.
Soliton solutions in noncommutative gauge theory were introduced by
Polychronakos in \cite{soliton}. The papers listed in \cite%
{soliton2,soliton3} contributed a lot of essential work to the study of
solitons in noncommutative gauge theory. The important finding of Gopakumar,
Minwalla and Strominger that a projector may correspond to a soliton in the
noncommutative field theory in paper \cite{strominger}, shows the
significance of studying projection operators in various noncommutative
space. Reiffel \cite{Rieffel} constructed the complete set of projection
operators on the noncommutative torus $T^{2}$. On the basis, Boca studied
the projection operators on noncommutative orbifold \cite{boca} having
obtained some beautiful results and the well-known example of projection
operator for the case of $T^{2}/Z_{4}$ in terms of the theta function.
Martinec and Moore in their important article deeply studied soliton
solutions on a wide variety of orbifolds, and the relation between physics
and mathematics in this area \cite{martinec}. Gopakumar, Headrick and
Spradlin showed rather a clear method to construct the multi-soliton
solution on noncommutative integral torus with generic $\tau $\cite{ghs}.
This approach can be used to construct the projection operators on the
integral noncommutative orbifold $T^{2}/Z_{N}$ \cite{yang}.

Some manifest covariant projectors with $Z_{4}$ symmetry on noncommutative
orbifold $T^{2}/Z_{4}$ were given \cite{boca}\cite{Deng}. In \cite{yang},we
have used the GHS construction to obtain a closed form for the projectors on
noncommutative orbifold $T^{2}/Z_{6}$ in terms of theta function. However,
Its form is complicated and not explicitly covariant. In this paper, by GHS
construction we give the projectors for integral $T^{2}/Z_{3}$ and $%
T^{2}/Z_{6}$, which is symmetric and manifestly covariant under $T^{2}/Z_{6}$
and $T^{2}/Z_{3}$ rotations . Also, the integration form of this expression
include all the projectors with minimal trace and continuous reduced
matrices with respect to the variables $k$ and $q$, just as that in \cite%
{Deng}.

This paper is organized as following: In Section 2, we briefly review the
operators on the noncommutative orbifold $T^{2}/Z_{N}$ and GHS construction.
In Section 3, we present the explicit and manifest covariant form for the
projectors on noncommutative orbifold $T^{2}/Z_{6}$. In the last section, we
provide the general covariant projection operators on noncommutative
orbifold integral $T^{2}/Z_{6}$ and $T^{2}/Z_{3}$. We conclude this paper
with some discussions.

\section{Noncommutative Orbifold $T^2/Z_N$}

In this section, we introduce operators on the noncommutative orbifold $%
T^{2}/Z_{N}$. Let two hermitian operators $\hat{y}_{1}$ and $\hat{y}_{2}$
satisfy the following commutation relation:

\begin{equation}
\lbrack \hat{y}_{1},\hat{y}_{2}]=i.
\end{equation}%
The operators constituted by the series of $\hat{y}_{1}$ and $\hat{y}_{2}$ 
\begin{equation}
\hat{O}=\sum_{m,n}C_{mn}\hat{y}_{1}^{m}\hat{y}_{2}^{n},~~~m,n\in \mathbb{Z}%
~and~m,n\geqslant 0
\end{equation}%
form a noncommutative plane $%
\mathbb{R}
^{2}$. All the operators in $%
\mathbb{R}
^{2}$ which commute with $U_{1}$ and $U_{2}$ defined by 
\begin{equation}
U_{1}=e^{-il\hat{y}_{2}},~~~~~~~~~~U_{2}=e^{il(\tau _{2}\hat{y}_{1}-\tau _{1}%
\hat{y}_{2})},
\end{equation}%
(where $l,\tau _{1},\tau _{2}$ are all real numbers and $l,\tau _{2}>0,\tau
=\tau _{1}+i\tau _{2}$), constitute the noncommutative torus $T^{2}$. We
have 
\begin{eqnarray}
U_{1}^{-1}\hat{y}_{1}U_{1} &=&\hat{y}_{1}+l,~~~~~~~U_{2}^{-1}\hat{y}%
_{1}U_{2}=\hat{y}_{1}+l\tau _{1},  \notag \\
U_{1}^{-1}\hat{y}_{2}U_{1} &=&\hat{y}_{2},~~~~~~~~~~~U_{2}^{-1}\hat{y}%
_{2}U_{2}=\hat{y}_{2}+l\tau _{2}.
\end{eqnarray}%
The operators $U_{1}$ and $U_{2}$ are two different wrapping operators
around the noncommutative torus and their commutation relation is $%
U_{1}U_{2}=U_{2}U_{1}e^{-2\pi i\frac{l^{2}\tau _{2}}{2\pi }}$. When $A=\frac{%
l^{2}\tau _{2}}{2\pi }$ is an integer, we have $[U_{1},U_{2}]=0$ and call
the noncommutative torus integral. Define two operators $u_{1}$ and $u_{2}$: 
\begin{eqnarray}
u_{1} &=&e^{-il\hat{y}_{2}/A},~~~~~~~~~~u_{2}=e^{-il(\tau _{2}\hat{y}%
_{1}-\tau _{1}\hat{y}_{2})/A},  \notag \\
u_{1}u_{2} &=&u_{2}u_{1}e^{2\pi
i/A},~~~~~~u_{1}^{A}=U_{1},~~u_{2}^{A}=U_{2}^{-1}.  \label{91}
\end{eqnarray}%
The operators on the noncommutative torus $T^{2}$ are composed of the
Laurent series of $u_{1}$ and $u_{2}$, 
\begin{equation}
\hat{O}_{T^{2}}=\sum_{m,n}C_{mn}^{\prime }u_{1}^{m}u_{2}^{n},  \label{eq:5}
\end{equation}%
where $m,n\in 
\mathbb{Z}
$ and $C_{00}^{\prime }$ is called the trace of the operator. Eq.(\ref{eq:5}%
) includes all the operators on the noncommutative torus $T^{2}$, satisfying
the invariant relation under action of $\left\{ U_{i}\right\} :U_{i}^{-1}%
\hat{O}_{T^{2}}U_{i}=\hat{O}_{T^{2}}$. We may rewrite the equation(\ref{eq:5}%
) as

\begin{equation}
\hat{O}_{T^{2}}=\sum_{s,t=0}^{A-1}u_{1}^{s}u_{2}^{t}\Psi
_{st}(u_{1}^{A},u_{2}^{A}),  \label{eq:6}
\end{equation}%
where $\Psi _{st}$ is the coefficient function of the Laurent series of
operators $u_{1}^{A}$ and $u_{2}^{A}$. We call this formula standard
expansion for the operator on the noncommutative torus $T^{2}$. The trace of
the operator is the constant term's coefficient of $\Psi _{00}$. Next we
introduce rotation $R$ in noncommutative space $%
\mathbb{R}
^{2},$ 
\begin{equation}
R(\theta )=e^{-i\theta \frac{\hat{y}_{1}^{2}+\hat{y}_{2}^{2}}{2}+i\frac{%
\theta }{2}}
\end{equation}%
with 
\begin{eqnarray}
R^{-1}\hat{y}_{1}R &=&\cos \theta \hat{y}_{1}+\sin \theta \hat{y}_{2}, \\
R^{-1}\hat{y}_{2}R &=&\cos \theta \hat{y}_{2}-\sin \theta \hat{y}_{1}.
\end{eqnarray}%
Assume $\tau =\tau _{1}+i\tau _{2}=e^{2\pi i/N}$, $\theta =2\pi /N(N\in 
\mathbb{Z}
)$. Define $R_{N}\equiv R(2\pi /N)$ . Then $U_{i}^{\prime }\equiv
R_{N}^{-1}U_{i}R_{N}$ can be expressed by monomial of $\left\{ U_{i}\right\} 
$ and their inverses for $A=2,3,4,6$. For these cases we may introduce the
orbifold $T^{2}/Z_{N}$\cite{boca,martinec}. We call the operators invariant
under rotation $R_{N}$ on the noncommutative torus as operators on
noncommutative orbifold $T^{2}/Z_{N}$. We can also realize these operators
in Fock space. Introduce 
\begin{equation}
a=\frac{\hat{y}_{2}-i\hat{y}_{1}}{\sqrt{2}},~~~~~~~~~a^{+}=\frac{\hat{y}%
_{2}+i\hat{y}_{1}}{\sqrt{2}},
\end{equation}%
then 
\begin{eqnarray}
\lbrack a,a^{+}] &=&1,  \notag \\
R &=&e^{-i\theta a^{+}a}.  \label{99}
\end{eqnarray}%
From the above discussion, we know that the operators $U_{1}$ and $U_{2}$
commute with each other on the integral torus $T^{2}$ when $A$ is an
integer. So we can introduce a complete set of their common eigenstates,
namely $\left\vert k,q\right\rangle $ representation \cite{ghs,21,20} 
\begin{equation}
\left\vert k,q\right\rangle =\sqrt{\frac{l}{2\pi }}e^{-i\tau _{1}\hat{y}%
_{2}^{2}/2\tau _{2}}\sum_{j}e^{ijkl}|q+jl\rangle ,  \label{86}
\end{equation}%
where the ket on the right is the $\hat{y}_{1}$ eigenstate. We have 
\begin{eqnarray}
U_{1}\left\vert k,q\right\rangle &=&e^{-ilk}\left\vert k,q\right\rangle
,~~~~~~~U_{2}\left\vert k,q\right\rangle =e^{il\tau _{2}q}\left\vert
k,q\right\rangle =e^{2\pi iqA/\l }\left\vert k,q\right\rangle ,  \label{eq:7}
\\
id &=&\int_{0}^{\frac{2\pi }{l}}dk\int_{0}^{l}dq\left\vert k,q\right\rangle
\langle k,q|.  \label{36}
\end{eqnarray}%
It also satisfies 
\begin{equation}
\left\vert k,q\right\rangle =|k+\frac{2\pi }{l},q\rangle
=e^{ilk}|k,q+l\rangle .  \label{89}
\end{equation}%
\begin{equation}
u_{1}\left\vert k,q\right\rangle =\left\vert k,q+\frac{l}{A}\right\rangle
,~~~~~~~u_{2}\left\vert k,q\right\rangle =e^{-il\tau _{2}q/A}\left\vert
k,q\right\rangle =e^{-2\pi iq/\l }\left\vert k,q\right\rangle  \label{87}
\end{equation}%
Consider Eq. (\ref{eq:6}), namely the standard expansion of operators on $%
T^{2}$ we have 
\begin{equation}
\Psi _{st}(u_{1}^{A},u_{2}^{A})\left\vert k,q\right\rangle =\Psi
_{st}(e^{-ilk},e^{-2\pi iqA/l})\left\vert k,q\right\rangle \equiv \tilde{\psi%
}_{st}(k,q)\left\vert k,q\right\rangle ,  \label{eq:15}
\end{equation}

where $\tilde{\psi}_{st}$ is function of the independent variables $k$ and $%
q $, called symbol function of $\Psi _{st}(u_{1}^{A},u_{2}^{A})$. From (\ref%
{eq:15}), we see that the function $\tilde{\psi}_{st}$ is invariant when $%
q\rightarrow q+l/A$ or $k\rightarrow k+2\pi /l$, 
\begin{equation}
\tilde{\psi}_{st}(k+\frac{2\pi m}{l},q)=\tilde{\psi}_{st}(k,q+\frac{ln}{A})=%
\tilde{\psi}_{st}(k,q).~~m,n\in 
\mathbb{Z}
\label{eq:16}
\end{equation}%
As long as the symbol function is obtained, the operator on the
noncommutative torus can be completely determined. Introducing new basis $%
|k,q_{0},n\rangle \equiv |k,q_{0}+\frac{ln}{A}\rangle ,k\in \lbrack 0,\frac{%
2\pi }{l}),q_{0}\in \lbrack 0,\frac{l}{A})$, we have from (\ref{36}) 
\begin{equation}
\sum_{n=0}^{A-1}\int_{0}^{\frac{2\pi }{l}}dk\int_{0}^{\frac{l}{A}%
}dq_{0}\left\vert k,q_{0}+\frac{ln}{A}\right\rangle \left\langle k,q_{0}+%
\frac{ln}{A}\right\vert =id.
\end{equation}%
From the above equation and (\ref{87})(\ref{eq:16}), we see that when any
power of the operators $u_{1}$ and $u_{2}$ acts on the $|k,q_{0}+\frac{ln}{A}%
\rangle $, the result can be expanded in the basis $|k,q_{0}+\frac{ln}{A}%
\rangle $ with the same $k,q_{0}$. So all the operators on the
noncommutative torus don't change $k$ and $q_{0}$. We have 
\begin{equation}
\hat{O}_{T^{2}}\left\vert k,q_{0}+\frac{ln}{A}\right\rangle =\sum_{n^{\prime
}}M^{O}(k,q_{0})_{n^{\prime }n}\left\vert k,q_{0}+\frac{ln^{\prime }}{A}%
\right\rangle .
\end{equation}%
Thus, for every $k$ and $q_{0}$ we get an $A\times A$ matrix, called reduced
matrix $M^{o}(k,q_{0})$. We have 
\begin{equation}
\hat{A}\hat{B}\left\vert k,q_{0}+\frac{ln}{A}\right\rangle =\sum_{n^{\prime
}}\left( M^{A}(k,q_{0})M^{B}(k,q_{0})\right) _{n^{\prime }n}\left\vert
k,q_{0}+\frac{ln^{\prime }}{A}\right\rangle .  \label{88}
\end{equation}%
For the projection operator on torus $T^{2}$, 
\begin{equation}
P\left\vert k,q_{0}+\frac{ln}{A}\right\rangle =\sum_{n^{\prime
}}M(k,q_{0})_{n^{\prime }n}\left\vert k,q_{0}+\frac{ln^{\prime }}{A}%
\right\rangle .  \label{eq:23}
\end{equation}%
It is easy to find the sufficient and necessary condition for $P^{2}=P$ from
(\ref{88}) \cite{yang} 
\begin{equation}
M(k,q_{0})^{2}=M(k,q_{0}).
\end{equation}%
When $T^{2}$ satisfies $Z_{N}$ symmetry, since after $R_{N}$ rotation $%
U_{i}^{\prime }$ can be expressed by monomial of $\{U_{i}\}$ and their
inverses, the state vector $R_{N}|k,q_{0}+\frac{ln}{A}\rangle $ is still the
common eigenstate of the operators $U_{1}$ and $U_{2}$. With the
completeness of $\{\left\vert k,q+\frac{ls}{A}\right\rangle \}$, and to
consider the eigenvalues of $U_{i}$ in the $kq$ representation, this vector
can be expanded in the basis $\{\left\vert k^{\prime },q^{\prime }+\frac{%
ls^{\prime }}{A}\right\rangle \}$ 
\begin{equation}
R_{N}|k,q_{0}+\frac{ln}{A}\rangle =\sum_{n^{\prime }}A(k,q_{0})_{n^{\prime
}n}|k^{\prime },q_{0}^{\prime }+\frac{ln^{\prime }}{A}\rangle ,
\label{eq:17}
\end{equation}%
where $k^{\prime }\in \lbrack 0,2\pi /l),q^{\prime }\in \lbrack 0,l/A)$ are
definite\cite{yang}. Equation (\ref{eq:17}) gives 
\begin{equation}
R_{N}^{-1}\left\vert k^{\prime },q_{0}^{\prime }+\frac{ln^{\prime }}{A}%
\right\rangle =\sum_{n"}A^{-1}(k,q_{0})_{n"n^{^{\prime }}}\left\vert k,q_{0}+%
\frac{ln"}{A}\right\rangle .  \label{eq:18}
\end{equation}%
We can get the expression for the relation between $k^{\prime }$,$%
q_{0}^{\prime }$ and $k$, $q_{0}$, The mapping $W:(k,q_{0})\longrightarrow
(k^{\prime },q_{0}^{\prime }),W^{N}=id$, is essentially a linear relation,
and area-preserving. By this fact and since $R_{N}$ is unitary, we conclude
that the matrix $A$ is an unitary matrix\cite{yang} 
\begin{equation}
A^{\ast }(k,q_{0})_{nn^{\prime }}=A^{-1}(k,q_{0})_{n^{\prime }n}.
\label{eq:19}
\end{equation}%
Since the projector on the noncommutative orbifold $T^{2}/Z_{N}$ satisfies $%
R_{N}^{-1}PR_{N}=P$, then from (\ref{eq:23})(\ref{eq:17})(\ref{eq:18}) one
obtains 
\begin{equation}
R_{N}^{-1}PR_{N}|k,q_{0}+\frac{ln}{A}\rangle
=\sum_{n"}[A^{-1}(k,q_{0})M(k^{\prime },q_{0}^{\prime
})A(k,q_{0})]_{n"n}|k,q_{0}+\frac{ln"}{A}\rangle ,  \label{eq:35}
\end{equation}%
which should be equal to : 
\begin{equation}
P|k,q_{0}+\frac{ln}{A}\rangle =\sum_{n"}M(k,q_{0})_{n"n}|k,q_{0}+\frac{ln"}{A%
}\rangle .
\end{equation}%
So, we have 
\begin{equation}
M(k^{\prime },q_{0}^{\prime })=A(k,q_{0})M(k,q_{0})A^{-1}(k,q_{0})
\label{eq:10}
\end{equation}%
Thus the sufficient and necessary condition for the redued matrix of a
projector on noncommutative orbifold $T^{2}/Z_{N}$ to satisfy is: 
\begin{eqnarray}
M(k,q_{0})^{2} &=&M(k,q_{0}),  \label{eq:24} \\
M(k^{\prime },q_{0}^{\prime }) &=&A(k,q_{0})M(k,q_{0})A^{-1}(k,q_{0}).
\label{37}
\end{eqnarray}%
Next we study the relation between the coefficient function $\tilde{\psi}%
_{st}(k,q)$ and reduced matrix $M(k,q_{0})$. Due to (\ref{87})(\ref{eq:15})(%
\ref{eq:16}) and (\ref{eq:23})we have 
\begin{eqnarray}
P|k,q_{0}+\frac{ln}{A}\rangle &=&\sum_{s,t}u_{1}^{s}u_{2}^{t}\Psi
_{st}(u_{1}^{A},u_{2}^{A})|k,q_{0}+\frac{ln}{A}\rangle  \notag \\
&=&\sum_{s,t}e^{-2\pi i(q_{0}/l+n/A)t}\tilde{\psi}_{st}(k,q_{0})|k,q_{0}+%
\frac{l(n+s)}{A}\rangle  \notag \\
&=&\sum_{n^{\prime }}M(k,q_{0})_{n^{\prime }n}|k,q_{0}+\frac{ln^{\prime }}{A}%
\rangle .
\end{eqnarray}%
From the periodic condition of $|k,q\rangle $ (\ref{89}), for $n+s<A$ case,
we have 
\begin{equation}
M(k,q_{0})_{n+s,n}=\sum_{t=0}^{A-1}e^{-2\pi i(q_{0}/l+n/A)t}\tilde{\psi}%
_{st}(k,q_{0}),
\end{equation}%
and for $n+s\geq A$ case, we have 
\begin{equation}
M(k,q_{0})_{n+s-A,n}=\sum_{t=0}^{A-1}e^{-2\pi i(q_{0}/l+n/A)t}\tilde{\psi}%
_{st}(k,q_{0})e^{-ilk}.
\end{equation}%
Setting 
\begin{equation}
M(k,q_{0})_{n+s,n}=M(k,q_{0})_{n+s-A,n}e^{ilk},  \label{eq:1}
\end{equation}%
we can uniformly write as: 
\begin{equation}
M(k,q_{0})_{n+s,n}=\sum_{t=0}^{A-1}e^{-2\pi i(q_{0}/l+n/A)t}\tilde{\psi}%
_{st}(k,q_{0})  \label{eq:2}
\end{equation}%
and have 
\begin{equation}
\tilde{\psi}_{st}(k,q_{0})=\frac{1}{A}\sum_{r=0}^{A-1}M(k,q_{0})_{r+s,r}e^{2%
\pi i(q_{0}/l+r/A)t}.  \label{eq:3}
\end{equation}%
Eq.(\ref{eq:2}) and (\ref{eq:3}) is the relation between $\tilde{\psi}_{st}$
and the elements of reduced matrix $M$.

We set (This is the $GHS$ construction)\cite{ghs} 
\begin{equation}
M(k,q_{0})_{nn^{\prime }}=\frac{\langle k,q_{0}+\frac{ln}{A}|\phi
_{1}\rangle \langle \phi _{2}|k,q_{0}+\frac{ln^{\prime }}{A}\rangle }{%
\sum_{n^{"}}\langle k,q_{0}+\frac{ln"}{A}|\phi _{1}\rangle \langle \phi
_{2}|k,q_{0}+\frac{ln"}{A}\rangle }.
\end{equation}%
It satisfies (\ref{eq:24}) and as long as $R|\phi _{j}\rangle =e^{i\alpha
_{j}}|\phi _{j}\rangle $, it also satisfies (\ref{37})\cite{Deng}. Notice
that this equation satisfies (\ref{eq:1}). We then have 
\begin{eqnarray}
\tilde{\psi}_{st}(k,q_{0}) &=&\frac{1}{A}%
\sum_{r=0}^{A-1}M(k,q_{0})_{r+s,r}e^{2\pi i(q_{0}/l+r/A)t}  \notag \\
&=&\frac{\frac{1}{A}\sum_{r=0}^{A-1}\langle k,q_{0}+\frac{l(r+s)}{A}|\phi
_{1}\rangle \langle \phi _{2}|k,q_{0}+\frac{lr^{\prime }}{A}\rangle e^{2\pi
i(q_{0}/l+r/A)t}}{\sum_{r}\langle k,q_{0}+\frac{lr}{A}|\phi _{1}\rangle
\langle \phi _{2}|k,q_{0}+\frac{lr}{A}\rangle }  \notag \\
&=&\frac{\tilde{f}_{st}(k,q_{0})}{A\tilde{f}_{00}(k,q_{0})},  \label{43}
\end{eqnarray}%
where 
\begin{equation}
\tilde{f}_{st}(k,q_{0})\equiv \sum_{r=0}^{A-1}\langle k,q_{0}+\frac{l(r+s)}{A%
}|\phi _{1}\rangle \langle \phi _{2}|k,q_{0}+\frac{lr}{A}\rangle e^{2\pi
i(q_{0}/l+r/A)t},
\end{equation}%
with 
\begin{eqnarray}
\tilde{f}_{st}(k,q_{0}) &=&\tilde{f}_{st}(k,q_{0}+l/A)=\tilde{f}_{st}(k+2\pi
/l,q_{0}), \\
\tilde{f}_{st}(k,q_{0}) &=&\tilde{f}_{s+A,t}(k,q_{0})e^{-ilk} \\
&=&\tilde{f}_{s,t+A}(k,q_{0})e^{-2\pi iq_{0}A/l}.
\end{eqnarray}%
Define 
\begin{eqnarray}
u &=&\frac{lk}{2\pi },~~~v=\frac{q_{0}}{l},  \label{92} \\
f_{st}(u,Av) &\equiv &\tilde{f}_{st}(k,q_{0}).  \label{60}
\end{eqnarray}%
So the function $f_{st}(u,Av)$ is function of the independent variables $u$
and $Av$ with period $1$. Similarly define 
\begin{equation}
\psi _{st}(u,Av)\equiv \tilde{\psi}_{st}(k,q_{0}),  \label{eq:12}
\end{equation}%
we have%
\begin{equation}
\psi _{st}(u,Av)=\frac{f_{st}(u,Av)}{Af_{00}(u,Av)}.  \label{97}
\end{equation}%
Let 
\begin{eqnarray*}
X &\equiv &e^{-ilk}=e^{-2\pi iu}, \\
Y &\equiv &e^{-2\pi iqA/l}=e^{-2\pi iAv}.
\end{eqnarray*}%
If we change the variables $X$ and $Y$ into $u_{1}^{A}$ and $u_{2}^{A}$
respectively in $\psi _{st}(u,Av),$ the standard form (\ref{eq:6}) of the
projection operator can be easily obtained. So the key question is to find
out $\tilde{f}_{st}(k,q_{0})$. For simplicity, we set 
\begin{equation}
|\phi _{1}\rangle =|\phi _{2}\rangle =|0\rangle ,~~~a|0\rangle
=0,~~~R_{N}|0\rangle =|0\rangle .
\end{equation}%
After some derivation, we have \cite{ghs} 
\begin{eqnarray}
&<&k,q|0>\equiv C_{0}(k,q)=\frac{1}{\sqrt{l}\pi ^{1/4}}\theta \left[ 
\begin{array}{l}
0 \\ 
0%
\end{array}%
\right] (\frac{q}{l}+\frac{\tau k}{l\tau _{2}},\frac{\tau }{A})e^{-\frac{%
\tau }{2i\tau _{2}}k^{2}+ikq} \\
&=&\sqrt{\frac{Ai}{l\tau \sqrt{\pi }}}\theta \left[ 
\begin{array}{l}
0 \\ 
0%
\end{array}%
\right] (\frac{lk}{2\pi }+\frac{Aq}{l\tau },-\frac{A}{\tau })e^{-\pi i\frac{%
Aq^{2}}{\tau l^{2}}},
\end{eqnarray}%
where 
\begin{equation}
\theta (z,\tau )\equiv \theta \left[ 
\begin{array}{l}
0 \\ 
0%
\end{array}%
\right] (z,\tau ),
\end{equation}%
\begin{equation}
\theta \left[ 
\begin{array}{c}
a \\ 
b%
\end{array}%
\right] (z,\tau )=\sum_{m}e^{\pi i\tau (m+a)^{2}}e^{2\pi i(m+a)(z+b)}.
\label{50}
\end{equation}%
Define%
\begin{equation}
g_{ss^{\prime }}(u,v)\equiv \langle k,q+\frac{ls}{A}|0\rangle \langle 0|k,q+%
\frac{ls^{\prime }}{A}\rangle .
\end{equation}%
Then we get for real $u$ and $v$,%
\begin{eqnarray}
f_{st}(u,Av) &=&\sum_{r=0}^{A-1}g_{s+r,r}(u,v)\times e^{2\pi it(\frac{r}{A}%
+v)}  \notag \\
&=&\sum_{r}\frac{1}{l\sqrt{\pi }}\theta (v+\frac{u\tau +s+r}{A},\frac{\tau }{%
A})\theta (v+\frac{u\tau ^{\ast }+r}{A},\frac{-\tau ^{\ast }}{A})\times
e^{2\pi it(v+\frac{r}{A})}  \notag \\
&&\times e^{\pi i\frac{\tau -\tau ^{\ast }}{A}u^{2}+2\pi i\frac{s}{A}u} 
\notag \\
&=&\frac{A}{l|\tau |\sqrt{\pi }}\sum_{r}\theta (u+\frac{A}{\tau }(v+\frac{s+r%
}{A}),-\frac{A}{\tau })\theta (u+\frac{A}{\tau ^{\ast }}(v+\frac{r}{A}),%
\frac{A}{\tau ^{\ast }})  \notag \\
&&\times e^{-\pi i\frac{A}{\tau }(v+\frac{s+r}{A})^{2}+\pi i\frac{A}{\tau
^{\ast }}(v+\frac{r}{A})^{2}}\times e^{2\pi it(v+\frac{r}{A})}.  \label{48}
\end{eqnarray}%
Then from (\ref{48}) and properties of theta functions,we have 
\begin{eqnarray}
f_{st}(u+1,Av) &=&f_{st}(u,Av+1)=f_{st}(u,Av), \\
f_{st}(u+A\tau ,Av) &=&e^{-2\pi i(2u+A(\tau +\tau ^{\ast })v+\frac{A}{2}%
(\tau -\tau ^{\ast })+\frac{s}{\tau })}f_{st}(u,Av), \\
f_{st}(u,A\tau +Av) &=&e^{-2\pi i(2Av+(\tau +\tau ^{\ast })u+A\frac{\tau
-\tau ^{\ast }}{2}-t\tau )}f_{st}(u,Av).
\end{eqnarray}%
These are the brief review of the $GHS$ construction the projection
operators on noncommutative orbifold $T^{2}/Z_{N}$ . In the next section, we
will concretely discuss how to construct the manifest covariant projectors
on noncommutative orbifold $T^{2}/Z_{6}.$

\section{The covariant projectors on noncommutative orbifold $T^{2}/Z_{6}$}

In the above section, we reviewed some results for projectors on
noncommutative orbifold $T^{2}/Z_{N}$. Boca and we presented some manifest
covariant projectors with $Z_{4}$ symmetry on noncommutative integral
orbifold $T^{2}/Z_{4}$\cite{boca}\cite{Deng}. In \cite{yang}, we have
presented a closed form for projectors on the noncommutative orbifold $%
T^{2}/Z_{6}$ in terms of elliptic function. However, its form is not
explicitly covariant. In this section, we are devoted to develop the
manifest covariant form for projectors on the noncommutative orbifold $%
T^{2}/Z_{6}$ by $GHS$ construction. In the case that $\tau =\tau _{6}=e^{%
\frac{\pi i}{3}},$ we have%
\begin{eqnarray}
f_{st}(u+1,Av) &=&f_{st}(u,Av+1)=f_{st}(u,Av),  \label{38} \\
f_{st}(u+A\tau ,Av) &=&e^{-2\pi i(2u+Av+A\tau -\frac{A}{2}+\frac{s}{\tau }%
)}f_{st}(u,Av),  \label{39} \\
f_{st}(u,A\tau +Av) &=&e^{-2\pi i(2Av+u+A\tau -\frac{A}{2}-t\tau
)}f_{st}(u,Av).  \label{40}
\end{eqnarray}%
From this, it can be proved that $f_{st}(u,Av)$ belongs to a
three-dimensional linear space. We can define the basis of this space as 
\begin{equation}
\theta (Av+\alpha )\theta (Av+u+\beta )\theta (u+\gamma )\equiv e(u,Av),
\label{90}
\end{equation}%
such that three linearly independent basis are enough to construct any
function satisfying conditions (\ref{38})-(\ref{40}). Here $\alpha ,\beta
,\gamma $ are parameters to be given later and we denote 
\begin{equation*}
\theta (z)\equiv \theta (z,A\tau )\equiv \theta \left[ 
\begin{array}{c}
0 \\ 
0%
\end{array}%
\right] (z,A\tau )
\end{equation*}%
for simplicity. (In the following, theta function without modular parameter
means its modular parameter is $A\tau $) We have from (\ref{90}) 
\begin{eqnarray}
e(u+1,Av) &=&e(u,Av+1)=e(u,Av), \\
e(u+A\tau ,Av) &=&e^{-2\pi i(2u+Av+A\tau +\beta +\gamma )}e(u,Av), \\
e(u,A\tau +Av) &=&e^{-2\pi i(u+2Av+A\tau +\alpha +\beta )}e(u,Av).
\end{eqnarray}%
Thus we require that 
\begin{equation}
\alpha +\beta =-\frac{A}{2}-t\tau ,~~\beta +\gamma =-\frac{A}{2}+\frac{s}{%
\tau },  \label{45}
\end{equation}%
where $\tau =e^{\frac{\pi i}{3}}.$ Next, we will consider the covariant
property for the projectors. From the definition, it is easy to get for $%
R=R_{6}$ 
\begin{eqnarray}
&&u_{1}^{\prime }=R^{-1}u_{1}R=u_{2}^{-1},~~~u_{2}^{\prime
}=R^{-1}u_{2}R=e^{-\pi i/A}u_{1}u_{2}, \\
&&~~~~~u_{1}u_{2}=e^{2\pi i/A}u_{2}u_{1}  \label{84}
\end{eqnarray}%
Define 
\begin{equation*}
c=e^{-\pi i/A},
\end{equation*}%
then 
\begin{equation}
R^{-1}PR=\sum_{st}u_{1}^{t}u_{2}^{t-s}c^{-2st+t^{2}}\Psi
_{st}(u_{2}^{-A},c^{A^{2}}u_{1}^{A}u_{2}^{A}).  \label{41}
\end{equation}%
We have from (\ref{eq:7})(\ref{87}) and (\ref{92}) 
\begin{eqnarray}
u_{1}^{A}\left\vert k,q\right\rangle  &=&e^{-2\pi iu}\left\vert
k,q\right\rangle ,~~~~u_{2}^{A}\left\vert k,q\right\rangle =e^{-2\pi
iAv}\left\vert k,q\right\rangle  \\
u_{1}^{\prime A}\left\vert k,q\right\rangle  &=&e^{-2\pi i(-Av)}\left\vert
k,q\right\rangle ,~~~~u_{2}^{\prime A}\left\vert k,q\right\rangle =e^{-\pi
iA}e^{-2\pi i(u+Av)}\left\vert k,q\right\rangle 
\end{eqnarray}%
From (\ref{eq:15}) (\ref{eq:12}) we have 
\begin{eqnarray}
\Psi _{st}(u_{1}^{A},u_{2}^{A})\left\vert k,q\right\rangle  &\equiv &\psi
_{st}(u,Av)\left\vert k,q\right\rangle   \notag \\
R^{-1}\Psi _{st}(u_{1}^{A},u_{2}^{A})R\left\vert k,q\right\rangle  &=&\Psi
_{st}(u_{1}^{\prime A},u_{2}^{\prime A})\left\vert k,q\right\rangle  \\
&=&\psi _{st}(-Av,-\frac{A}{2}+u+Av)\left\vert k,q\right\rangle .
\end{eqnarray}%
That is, the variables $u$ and $Av$ change as 
\begin{equation}
u\rightarrow -Av,Av\rightarrow -\frac{A}{2}+u+Av  \label{53}
\end{equation}%
under the rotation $R=R_{6}.$ Therefore, when $P=R^{-1}PR$, the formulation (%
\ref{eq:6}) and (\ref{41}) demand%
\begin{equation}
c^{-2st+t^{2}}\psi _{st}(-Av,-\frac{A}{2}+u+Av)=\psi _{t,t-s}(u,Av).
\label{44}
\end{equation}%
Notice that $\psi _{00}(u,Av)$\quad is invariant under rotation $R$. From (%
\ref{43}), we can get 
\begin{equation}
\tilde{\psi}_{st}(k,q_{0})\equiv \psi _{st}(u,Av)=\frac{f_{st}(u,Av)}{%
Af_{00}(u,Av)}.  \label{108}
\end{equation}%
If we construct $f_{st}(u,Av)$ satisfying the relation similar to (\ref{44})
obtaining $\psi _{st}(u,Av)$ by (\ref{108}), and set\quad $\psi
_{st}(u,Av)\left\vert kq\right\rangle =\Psi
_{st}(u_{1}^{A},u_{2}^{A})\left\vert k,q\right\rangle $,(see the text after
equation (\ref{97}))\quad then we can get the projector $P=\sum%
\limits_{st}u_{1}^{s}u_{2}^{t}\Psi _{st}(u_{1}^{A},u_{2}^{A})$ which is
invariant under rotation $R$. In the following, we wish to find a set of
covariant basis to construct such $f_{st}$. We write the basis as 
\begin{equation*}
e(u,Av)=\theta (Av+\alpha )\theta (Av+u+\beta )\theta (u+\gamma ).
\end{equation*}%
After rotation $R,$ it is turned into 
\begin{eqnarray}
e^{\prime }(u,Av) &=&\theta (-\frac{A}{2}+u+Av+\alpha )\theta (-\frac{A}{2}%
+u+\beta )\theta (Av-\gamma ) \\
&=&\theta (Av+u+\beta ^{\prime })\theta (u+\gamma ^{\prime })\theta
(Av+\alpha ^{\prime }).  \notag
\end{eqnarray}%
Thus the base vector change its parameters under the rotation as 
\begin{equation}
\alpha ^{\prime }=-\gamma ,~~\beta ^{\prime }=\alpha -\frac{A}{2},~~\gamma
^{\prime }=-\frac{A}{2}+\beta ~~(mod%
\mathbb{Z}
).  \label{46}
\end{equation}%
Now we take the transformation under the light of (\ref{44}) 
\begin{equation*}
s^{\prime }=t,~~t^{\prime }=t-s\Rightarrow t=s^{\prime }~~,s=s^{\prime
}-t^{\prime }
\end{equation*}%
The covariant base should satisfy 
\begin{equation}
e_{st}(-Av,-\frac{A}{2}+u+Av)=e_{t,t-s}(u,Av)=e_{s^{\prime },t^{\prime
}}(u,Av).
\end{equation}%
We set 
\begin{eqnarray}
\alpha  &=&\alpha _{st}=\alpha _{1}s+\alpha _{2}t+\alpha _{3},  \notag \\
\beta  &=&\beta _{st}=\beta _{1}s+\beta _{2}t+\beta _{3},  \notag \\
\gamma  &=&\gamma _{st}=\gamma _{1}s+\gamma _{2}t+\gamma _{3},  \label{51}
\end{eqnarray}%
here%
\begin{eqnarray}
\alpha _{1} &=&\frac{1}{2}+\frac{\sqrt{3}}{6}i,~\alpha _{2}=-\frac{\sqrt{3}}{%
3}i,~\alpha _{3}=\frac{B}{2},  \notag \\
\beta _{1} &=&\frac{1}{2}-\frac{\sqrt{3}}{6}i,~\beta _{2}=-\frac{1}{2}-\frac{%
\sqrt{3}}{6}i,~\beta _{3}=-\frac{A}{2}-\frac{B}{2}  \notag \\
\gamma _{1} &=&-\frac{\sqrt{3}}{3}i,~\gamma _{2}=-\frac{1}{2}+\frac{\sqrt{3}%
}{6}i,~\gamma _{3}=\frac{B}{2}.~~B\in \mathbb{Z}  \label{49}
\end{eqnarray}%
Then (\ref{45}) is satisfied. From (\ref{46})(\ref{51}) the variables $%
\alpha ,\beta ,\gamma $ transform as 
\begin{eqnarray}
\alpha ^{\prime } &=&\alpha _{1}s^{\prime }+\alpha _{2}t^{\prime }+\alpha
_{3}-B=\alpha _{s^{\prime }t^{\prime }}, \\
\beta ^{\prime } &=&\beta _{1}s^{\prime }+\beta _{2}t^{\prime }+\beta
_{3}+A=\beta _{s^{\prime }t^{\prime }},  \notag \\
\gamma ^{\prime } &=&\gamma _{1}s^{\prime }+\gamma _{2}t^{\prime }+\gamma
_{3}-B=\gamma _{s^{\prime }t^{\prime }}~~(mod%
\mathbb{Z}
).  \notag
\end{eqnarray}%
Then we have

\begin{equation}
e_{st}(-Av,-\frac{A}{2}+u+Av)=e_{t,t-s}(u,Av)=e_{s^{^{\prime }},t^{^{\prime
}}}(u,Av).  \label{47}
\end{equation}%
It really satisfies the covariant condition. Setting $B=0,1$ in (\ref{49}),
we obtain two linearly independent basis $e_{0}$ and $e_{1}$ which have the
covariant property (\ref{47}). We verify that $f_{st}(u,Av)$ of (\ref{48})
can be expanded by just the two basis in the following. We rewrite the $%
f_{st}(u,Av)$ in (\ref{48}) by taking 
\begin{equation}
\tilde{v}_{0}+\frac{\tilde{s}_{0}}{A}=v+\frac{\tau u}{A}+\frac{s}{A},~~%
\tilde{v}_{0}=v+\frac{\tau ^{\ast }u}{A}.  \label{80}
\end{equation}%
When $\tau =\tau _{6}=e^{\frac{2\pi }{6}i},$%
\begin{equation}
\frac{\tilde{s}_{0}}{A}=\frac{s}{A}+\frac{(2\tau -1)u}{A}~.  \label{81}
\end{equation}%
We get $f_{st}(u,Av)$ as follows: 
\begin{eqnarray}
f_{st}(u,Av) &=&\sum\limits_{r}\frac{1}{l\sqrt{\pi }}\theta (\tilde{v}_{0}+%
\frac{\tilde{s}_{0}+r}{A},\frac{\tau }{A})\theta (\tilde{v}_{0}+\frac{r}{A},-%
\frac{\tau ^{\ast }}{A})\times e^{2\pi itr/A}  \notag \\
&&\times e^{\pi i\frac{\tau -\tau ^{\ast }}{A}u^{2}+2\pi i\frac{s}{A}u+2\pi
itv}.  \label{61}
\end{eqnarray}%
Expanding the two theta functions involved in (\ref{61}) by (\ref{50}) we
obtain the following form. (see Appendix A for details) 
\begin{equation}
f_{st}(u,Av)=\frac{A}{l\sqrt{\pi }}\sum_{\delta =0,1}e^{2\pi i\phi
^{^{\prime }}}\theta \left( z^{\prime },\frac{2\tau -1}{A}\right) \theta
\left( w,A\left( 2\tau -1\right) \right) ,  \label{69}
\end{equation}%
where 
\begin{eqnarray}
z^{\prime } &=&-\delta \tau -\frac{\tau }{A}t+\frac{s}{A}+\frac{2\tau -1}{A}%
u,  \notag \\
w &=&(\delta A-t)\tau +s+2Av+u,  \notag \\
\phi ^{^{\prime }} &=&\delta A\frac{\tau -1}{2}+\delta (Av-(\tau -1)u)+\frac{%
2\tau -1}{2A}u^{2}+\frac{s}{A}u,  \notag \\
&&-\frac{\tau }{A}tu-\frac{st}{A}+\frac{\tau }{2A}t^{2}.
\end{eqnarray}%
From (\ref{48}) and (\ref{38})-(\ref{40}), the function $\tilde{f}%
_{st}\left( k,q\right) =f_{st}(u,Av)$ belongs to a three-dimensional space
spanned by functions of $u$ and $Av,$ $f_{st}(u,Av)$ can be expanded by the
following basis, 
\begin{eqnarray}
e_{0}\left( u,Av\right)  &=&\theta \left( Av+\alpha \right) \theta \left(
Av+u+\beta \right) \theta \left( u+\gamma \right) , \\
e_{1}\left( u,Av\right)  &=&\theta \left( Av+\alpha +\frac{1}{2}\right)
\theta \left( Av+u+\beta -\frac{1}{2}\right) \theta \left( u+\gamma +\frac{1%
}{2}\right) ,  \notag \\
e_{2}\left( u,Av\right)  &=&\theta \left( Av+\alpha +x\right) \theta \left(
Av+u+\beta -x\right) \theta \left( u+\gamma +x\right) ~~~~0<x\ll 1
\label{94}
\end{eqnarray}%
where $\alpha ,\beta ,\gamma $ are given in (\ref{51})(\ref{49}) with $B=0.$%
We have%
\begin{equation}
f_{st}=c_{0}e_{0}+c_{1}e_{1}+c_{2}e_{2}  \label{54}
\end{equation}%
For convenience of derivation we change the arguments as follows:%
\begin{eqnarray}
Av &=&\lambda -\alpha +a,~~~  \label{62} \\
u &=&-\gamma +b,  \label{63}
\end{eqnarray}

where$~\lambda =\frac{1}{2}\left( A\tau +1\right) .$ Notice $\beta -\alpha
-\gamma =\frac{A}{2}$ in the setting of (\ref{51})(\ref{49}) for $B=0$. Then
we have 
\begin{eqnarray}
e_{0} &=&\theta \left( \lambda +a\right) \theta \left( \lambda -\frac{A}{2}%
+a+b\right) \theta \left( b\right) ,  \notag \\
e_{1} &=&\theta \left( \lambda +a+\frac{1}{2}\right) \theta \left( \lambda -%
\frac{A}{2}+a+b-\frac{1}{2}\right) \theta \left( b+\frac{1}{2}\right) , 
\notag \\
e_{2} &=&\theta \left( \lambda +a+x\right) \theta \left( \lambda -\frac{A}{2}%
+a+b-x\right) \theta \left( b+x\right) .
\end{eqnarray}%
Based on the replacement of arguments given by (\ref{62}) and (\ref{63}), we
rewrite $f_{st}$ in (\ref{69}) by variables $a$ and $b,$ 
\begin{eqnarray}
f_{st}(u,Av)|_{v=\frac{\lambda -\alpha +a}{A}}^{u=-\gamma +b} &=&\frac{A}{l%
\sqrt{\pi }}e^{2\pi i\phi _{st}}\sum_{\delta =0,~1}e^{2\pi i\phi _{\delta
}\left( a,b\right) }\theta \left( \delta \left( 1-\tau \right) +\frac{2\tau
-1}{A}b,\frac{2\tau -1}{A}\right)  \notag \\
&&\times \theta \left( \delta A\tau +2\lambda +2a+b,A\left( 2\tau -1\right)
\right) ,  \label{55}
\end{eqnarray}%
where 
\begin{eqnarray*}
\phi _{st} &=&\frac{\sqrt{3}i}{6A}\left( s^{2}+t^{2}-st\right) -\frac{1}{2A}%
st \\
\phi _{\delta }\left( a,b\right) &=&\frac{\sqrt{3}i}{2A}b^{2}+\delta \left(
\delta A\frac{\tau -1}{2}+\lambda +a+\left( 1-\tau \right) b\right)
\end{eqnarray*}

\qquad\ 

In order to verify $c_{2}=0$ and determine the coefficients $c_{0},c_{1}$in (%
\ref{54}), consider the case of $b_{m}=\frac{1-A}{2}+\frac{mA}{2\tau -1},$ $%
m\in 
\mathbb{Z}
$ and $a=a_{n}=\frac{n}{2},n\in 
\mathbb{Z}
$ $.$ Since $\theta \left( n_{1}+\frac{1}{2}+\left( n_{2}+\frac{1}{2}\right)
\tau ,\tau \right) =0,n_{1},n_{2}\in 
\mathbb{Z}
,$ when $\delta =1,$ the first theta function in (\ref{55}) $\theta \left( 
\frac{\left( 1-2A\right) \left( 2\tau -1\right) }{2A}+m+\frac{1}{2},\frac{%
2\tau -1}{A}\right) $ vanishes. Therefore we have for $\tau =e^{\pi i/3}.$

\begin{eqnarray}
f_{st}\left( u,Av\right) |_{v=\frac{1}{A}(\lambda -\alpha
+a_{n})}^{u=-\gamma +b_{m}} &\equiv &f_{mst}  \notag \\
&=&\frac{A}{l\sqrt{\pi }}e^{2\pi i\phi _{st}+2\pi i\phi _{\delta }\mid
_{_{\delta =0}}}\theta \left( \frac{\left( 1-A\right) \left( 2\tau -1\right) 
}{2A}+m,\frac{2\tau -1}{A}\right)  \notag \\
&&\times \theta \left( A\tau +1+2a_{n}+\frac{1-A}{2}+\frac{mA}{2\tau -1}%
,A\left( 2\tau -1\right) \right)  \notag \\
&=&\frac{A}{l\sqrt{\pi }}e^{2\pi i\phi _{st}+2\pi i\phi _{\delta }\mid
_{_{\delta =0}}}\theta \left( \frac{\left( 1-A\right) \left( 2\tau -1\right) 
}{2A}+m,\frac{2\tau -1}{A}\right)  \notag \\
&&\times \theta \left( \frac{A\left( 2\tau -1\right) }{2}+\frac{1}{2}-\frac{m%
}{3}A(2\tau -1)+n,A\left( 2\tau -1\right) \right)  \label{103}
\end{eqnarray}%
for $u=-\gamma _{st}+b_{m},v=\frac{1}{A}(\lambda -\alpha _{st}+a_{n}),$
which is independent of $n\in 
\mathbb{Z}
.$

when $m=3p,$ $p\in 
\mathbb{Z}
,$ the second theta function in the right-hand side of (\ref{103}) vanishes, 
\begin{equation}
f_{mst}=0,  \label{96}
\end{equation}

when $m\neq 3p,$ define%
\begin{equation}
f_{st}(u,Av)=f_{mst}\neq 0.  \label{52}
\end{equation}

Next we check $e_{0},e_{1}$ and $e_{2}$ in these cases, obtaining

\begin{itemize}
\item If $a=0,$then $e_{0}=0$ and%
\begin{equation*}
e_{1}^{0}(m)\equiv e_{1}|_{a=0}=\theta (\lambda +\frac{1}{2})\theta (\lambda
-\frac{m}{3}A(2\tau -1))\theta (b_{m}+\frac{1}{2});
\end{equation*}

\item If $a=-\frac{1}{2},$ then $e_{1}=0$ and 
\begin{equation}
e_{0}^{0}(m)\equiv e_{0}|_{a=-\frac{1}{2}}=\theta (\lambda -\frac{1}{2}%
)\theta (\lambda -\frac{m}{3}A(2\tau -1))\theta (b_{m});
\end{equation}

\item When $m=3p,e_{0}=e_{1}=0,e_{2}\neq 0$.
\end{itemize}

From (\ref{54}), (\ref{96})we can obtain 
\begin{equation}
c_{2}=0  \label{93}
\end{equation}%
and 
\begin{equation}
f_{st}(u,Av)=c_{0}e_{0}+c_{1}e_{1}.  \label{95}
\end{equation}%
Then we take $m\neq 3p$, giving 
\begin{eqnarray*}
f_{mst} &=&c_{1}e_{1}^{0}(m) \\
f_{mst} &=&c_{0}e_{0}^{0}(m)
\end{eqnarray*}%
\begin{equation}
\implies c_{1}=\frac{f_{mst}}{e_{1}^{0}(m)},c_{0}=\frac{f_{mst}}{e_{0}^{0}(m)%
},  \label{106}
\end{equation}%
So we have 
\begin{equation}
f_{st}=f_{mst}(\frac{e_{0}}{e_{0}^{0}(m)}+\frac{e_{1}}{e_{1}^{0}(m)}).
\label{57}
\end{equation}%
Besides, we have%
\begin{equation*}
\frac{f_{mst}}{f_{m00}}=e^{2\pi i\phi _{st}},
\end{equation*}%
and the ratio%
\begin{equation}
\frac{e_{0}^{0}(m)}{e_{1}^{0}(m)}=\frac{\theta (b_{m})}{\theta (b_{m}+\frac{1%
}{2})}  \label{58}
\end{equation}%
doesn't contain $s$ and $t.$ Thus from  (\ref{97})(\ref{57}) and (\ref{58})
we have 
\begin{equation}
\psi _{st}(u,Av)=\frac{f_{st}}{Af_{00}}=\frac{e^{2\pi i\phi _{st}}}{A}\frac{%
\theta (b_{m}+\frac{1}{2})e_{0}(st)+\theta (b_{m})e_{1}(st)}{\theta (b_{m}+%
\frac{1}{2})e_{0}(00)+\theta (b_{m})e_{1}(00)},  \label{98}
\end{equation}%
where 
\begin{equation*}
e_{j}(st)=\theta (Av+\alpha _{st}+\frac{1}{2}j)\theta (Av+u+\beta _{st}+%
\frac{1}{2}j)\theta (u+\gamma _{st}+\frac{1}{2}j).
\end{equation*}%
Let%
\begin{equation*}
\Theta (e^{-2\pi ix})\equiv \theta (x,A\tau ).
\end{equation*}%
From (\ref{eq:6}) and (\ref{eq:15})%
\begin{equation}
P_{Z_{6}}=\sum_{s,t=0}^{A-1}u_{1}^{s}u_{2}^{t}e^{2\pi i\phi _{st}}\frac{%
\theta (b_{m}+\frac{1}{2})\varepsilon _{0}(st)+\theta (b_{m})\varepsilon
_{1}(st)}{A[\theta (b_{m}+\frac{1}{2})\varepsilon _{0}(00)+\theta
(b_{m})\varepsilon _{1}(00)]},  \label{100}
\end{equation}%
where%
\begin{eqnarray}
\varepsilon _{j}(st) &=&\theta (l(\tau _{2}\hat{y}_{1}-\tau _{1}\hat{y}%
_{2})+\alpha _{st}+\frac{i}{2},A\tau )\times \theta (l(\left( \tau
_{2}+1\right) \hat{y}_{1}-\tau _{1}\hat{y}_{2})+\beta _{st}+\frac{j+A}{2}%
,A\tau )  \notag \\
&&\times \theta (l\hat{y}_{2}+\gamma _{st}+\frac{j}{2},A\tau )  \notag \\
&=&\Theta (u_{1}^{A}e^{2\pi i\left( \alpha _{st}+\frac{1}{2}j\right)
})\times \Theta (u_{2}^{A}e^{2\pi i\left( \alpha _{st}+\frac{1}{2}j\right)
})\times \Theta (u_{1}^{A}u_{2}^{A}e^{2\pi i\left( \gamma _{st}+\frac{1}{2}%
j\right) })  \label{59}
\end{eqnarray}%
Note that $\frac{A}{2}$ included in the second $\theta $ function in $%
\varepsilon _{j}(st)$ arise from 
\begin{equation*}
\lbrack l(\tau _{2}\hat{y}_{1}-\tau _{1}\hat{y}_{2}),l\hat{y}_{2}]\equiv
\lbrack A\hat{v},\hat{u}]=il^{2}\tau _{2}=2\pi iA,
\end{equation*}%
\begin{equation*}
e^{A\hat{v}+\hat{u}}=e^{A\hat{v}}e^{\hat{u}}e^{-\pi iA}=e^{\hat{u}}e^{A\hat{v%
}}e^{\pi iA},
\end{equation*}%
\begin{eqnarray*}
e^{A\hat{v}}e^{\hat{u}}\left\vert kq\right\rangle 
&=&u_{1}^{A}u_{2}^{A}\left\vert kq\right\rangle =e^{Av+u}\left\vert
kq\right\rangle  \\
&=&e^{A\hat{v}+\hat{u}+\pi iA}\left\vert kq\right\rangle .
\end{eqnarray*}%
In (\ref{100}),(\ref{59}), the parameters $\alpha _{st},\beta _{st},\gamma
_{st}$ are given by (\ref{51}) and (\ref{49}) with $B=0$,%
\begin{eqnarray}
\alpha _{st} &=&[e^{\frac{\pi i}{6}}s+e^{-\frac{\pi i}{2}}t]\frac{\sqrt{3}}{3%
}, \\
\beta _{st} &=&e^{-\frac{\pi i}{3}}\alpha _{st}-\frac{A}{2}, \\
\gamma _{st} &=&e^{-\frac{2\pi i}{3}}\alpha _{st}, \\
\phi _{st} &=&\frac{\sqrt{3i}}{6A}(s^{2}+t^{2}-st)-\frac{st}{2A}, \\
b_{m} &=&\frac{1-A}{2}+m\frac{A}{2\tau -1}=\frac{1-A}{2}-\frac{m}{3}A(2\tau
-1),~m\neq 3p.
\end{eqnarray}%
We take $m=3p+M,~M=\pm 1$ to obtain 
\begin{equation}
\frac{\theta (b_{m}+\frac{1}{2})}{\theta (b_{m})}=\frac{\theta (\frac{A}{2}-%
\frac{A(2\tau -1)}{3},A\tau )}{\theta (\frac{A}{2}-\frac{1}{2}-\frac{A(2\tau
-1)}{3},A\tau )},
\end{equation}%
which is independent of the choice of $M,p.$ Now we check the covariance
under transformation $R$. In the following, we find the expression (\ref{100}%
) possesses manifest covariance. Actually, $e_{0}$ and $e_{1}$ are the
covariant functions obtained from (\ref{49}) by taking $B=0$ and $B=1.$
Therefore they satisfy the covariant relation (\ref{47}). We then check(\ref%
{44}). From (\ref{98}), the exponent of phase factor related to $st$ on the
left-handed side of (\ref{44}) is proportional to 
\begin{equation*}
\phi _{st}-\frac{1}{2A}\left( t^{2}-2st\right) =\frac{\sqrt{3}i}{6A}\left(
s^{2}+t^{2}-st\right) -\frac{1}{2A}\left( t^{2}-st\right) 
\end{equation*}%
On the right-handed side phase factor is the exponent of 
\begin{eqnarray*}
&&\frac{\sqrt{3}i}{6A}\left( \left( t-s\right) ^{2}+t^{2}-t\left( t-s\right)
\right) -\frac{1}{2A}\left( t^{2}-st\right)  \\
&=&\frac{\sqrt{3}i}{6A}\left( s^{2}+t^{2}-st\right) -\frac{1}{2A}\left(
t^{2}-st\right) 
\end{eqnarray*}%
The two phase factors equal. From (\ref{47})(\ref{59}), we know that the $%
\psi _{st}$ given by (\ref{98}) really satisfies the covariance relation (%
\ref{44}). So (\ref{100}) is the solution of projector which possesses
manifest rotational covariance.

Now we have constructed the explicit and Manifestly covariant form for the
projection operators on noncommutative integral orbifold $T^{2}/Z_{6}$ with
trace $\frac{1}{A}$.

\section{The general covariant projection operators}

In this section we construct the general projectors with manifest covariant
property by GHS construction. Instead of the vacuum $\left\vert
0\right\rangle ,$ we take 
\begin{equation}
|\phi _{j}\rangle =\int d^{2}zF_{j}(z)|z\rangle ,~~j=1,2
\end{equation}%
where $|z\rangle $ is the coherent state satisfying the relation $a|z\rangle
=\frac{l}{i\sqrt{2}}z|z\rangle ,$ $F_{j}(z)$ is an arbitrary continuous
function of the argument $z$. Then for $R$ in (\ref{99}), we have 
\begin{equation}
R|z\rangle =|e^{-i\theta }z\rangle  \label{25.2}
\end{equation}%
Now take $\theta =\frac{\pi }{3},R=R_{6}.$ When $F_{j}(z)$ satisfies the $%
Z_{6}$ symmetry 
\begin{equation}
F_{j}(e^{\frac{\pi i}{3}}z)=e^{i\alpha _{j}}F_{j}(z),  \label{25.4}
\end{equation}%
we have 
\begin{equation}
R|\phi _{j}\rangle =e^{i\alpha _{j}}|\phi _{j}\rangle .
\end{equation}%
Then we may obtain a projector in $T^{2}/Z_{6}$ from (\ref{43}). In this
case, we have%
\begin{equation}
f_{st}(u,Av)=\int d^{2}z_{1}d^{2}z_{2}\sum_{r=0}^{A-1}\langle k,q_{0}+\frac{%
l(r+s)}{A}|z_{1}\rangle \langle z_{2}|k,q_{0}+\frac{lr}{A}\rangle e^{2\pi i(%
\frac{q_{0}}{l}+\frac{r}{A})t}F_{1}(z_{1})F_{2}(z_{2})^{\ast }.
\end{equation}%
Define%
\begin{equation*}
G(u,v,z_{1},z_{2}^{\ast })_{ss^{\prime }}\equiv \langle k,q_{0}+\frac{ls}{A}%
|z_{1}\rangle \langle z_{2}|k,q_{0}+\frac{ls^{\prime }}{A}\rangle .
\end{equation*}%
We have 
\begin{equation*}
f_{st}(u,Av)\equiv \int d^{2}z_{1}d^{2}z_{2}F_{1}(z_{1})F_{2}(z_{2})^{\ast
}f_{st}(u,Av,z_{1},z_{2}^{\ast }),
\end{equation*}%
where 
\begin{eqnarray}
f_{st}(u,Av,z_{1},z_{2}^{\ast }) &=&\sum_{r=0}^{A-1}G(u,v,z_{1},z_{2}^{\ast
})_{s+r,r}e^{2\pi it(\frac{r}{A}+v)}  \notag \\
&=&[c_{0}\theta (Av+\alpha -Az_{1}\alpha _{1}-Az_{2}^{\ast }\beta
_{1})\theta (Av+u+\beta -Az_{1}\beta _{1}-Az_{2}^{\ast }\alpha _{1})  \notag
\\
&&\times \theta (u+\gamma -Az_{1}\gamma _{1}+Az_{2}^{\ast }\gamma
_{1})+c_{1}\theta (Av+\alpha -Az_{1}\alpha _{1}-Az_{2}^{\ast }\beta _{1}+%
\frac{1}{2})  \notag \\
&&\times \theta (Av+u+\beta -Az_{1}\beta _{1}-Az_{2}^{\ast }\alpha _{1}-%
\frac{1}{2})\theta (u+\gamma -Az_{1}\gamma _{1}+Az_{2}^{\ast }\gamma _{1}+%
\frac{1}{2})]  \notag \\
&&\times e^{2\pi is(z_{1}-z_{2}^{\ast })\gamma _{1}+2\pi it(z_{1}\alpha
_{1}+z_{2}^{\ast }\beta _{1})+\frac{l^{2}}{4}(2z_{1}z_{2}^{\ast
}+|z_{1}|^{2}+\left\vert z_{2}\right\vert ^{2})}.
\end{eqnarray}%
The proof is given in Appendix B.

Due to (\ref{43}), one has%
\begin{equation}
\psi _{st}(u,Av)=\frac{\int
d^{2}z_{1}d^{2}z_{2}f_{st}(u,Av,z_{1},z_{2}^{\ast
})F_{1}(z_{1})F_{2}(z_{2})^{\ast }}{A\int
d^{2}z_{1}d^{2}z_{2}f_{00}(u,Av,z_{1},z_{2}^{\ast
})F_{1}(z_{1})F_{2}(z_{2})^{\ast }},  \label{111}
\end{equation}%
Therefore we obtain the explicit form for the general projection operators
as follows, 
\begin{equation}
P=\frac{1}{A}\sum_{s,t=0}^{A-1}u_{1}^{s}u_{2}^{t}e^{2\pi i\phi _{st}}\frac{%
\theta (b_{m}+\frac{1}{2})\epsilon _{0}(st)+\theta (b_{m})\epsilon _{1}(st)}{%
\theta (b_{m}+\frac{1}{2})\epsilon _{0}(00)+\theta (b_{m})\epsilon _{1}(00)},
\label{26.2}
\end{equation}%
where 
\begin{equation*}
\epsilon _{j}(s,t)=\int dz_{1}dz_{2}F_{1}(z_{1})F_{2}(z_{2})^{\ast
}E_{j}(s,t,z_{1},z_{2}^{\ast }),~~~~
\end{equation*}%
and 
\begin{eqnarray}
E_{j}(s,t,z_{1},z_{2}^{\ast }) &=&\Theta \left( u_{1}^{A}e^{2\pi i(\alpha
_{st}+\frac{j}{2})}e^{-2\pi iA(z_{1}\alpha _{1}+z_{2}^{\ast }\beta
_{1})}\right) \times \Theta \left( u_{1}^{A}u_{2}^{A}e^{2\pi i(\beta _{st}+%
\frac{j}{2})}e^{-2\pi iA(z_{1}\beta _{1}+z_{2}^{\ast }\alpha _{1})}\right)  
\notag \\
&&\times \Theta \left( u_{2}^{A}e^{2\pi i(\gamma _{st}+\frac{j}{2})}e^{-2\pi
iA(z_{1}-z_{2}^{\ast })\gamma _{1}}\right) .
\end{eqnarray}%
When $F_{j}(z_{j})$ satisfies $Z_{6}$ symmetry, namely in the case $\theta =%
\frac{\pi }{3}$ in formula (\ref{25.2}), (\ref{26.2}) shows the projectors $%
P_{Z_{6}}$, obviously the projector also belongs to $P_{Z_{3}}$. Just as
proved in \cite{Deng}, we have gotten all the projectors with trace $\frac{1%
}{A}$ on the orbifolds $T^{2}/Z_{6}$ including the case that $\tilde{\psi}%
_{st}(k,q_{0})$ is an analytic function. When $F_{j}(z_{j})$ satisfies $Z_{3}
$ symmetry but does not satisfy $Z_{6}$ symmetry, namely%
\begin{equation}
F_{j}(e^{\frac{2\pi i}{3}}z)=e^{i\alpha _{j}}F_{j}(z),
\end{equation}%
\begin{equation}
F_{j}(e^{\frac{\pi i}{3}}z)\neq const.F_{j}(z),
\end{equation}%
then (\ref{26.2}) gives a projector of $T^{2}/Z_{3},$ but it is not a
projector of $T^{2}/Z_{6}.$ It is shown that the form of our solution
possesses manifest covariance under rotation in Appendix B.

\section{Discussion}

We have found the complete set of projectors in analytic form with trace $%
\frac{1}{A}$ in all the cases of integral orbifold $T^{2}/Z_{N}$.( in the
case of $T^{2}/Z_{4}$ refer to \cite{Deng} ), of course the case with trace $%
\frac{A-1}{A}$ is naturally obtained via the case with trace $\frac{1}{A}$
by $P^{\prime }=id-P.$However we haven't obtained analytic solutions about
projectors with an arbitrary trace $\frac{A-m}{A},1<m<A-1$, which is an
intriguing question that is closely related to the resolvent of the case
that $A$ is a rational number but not integer number. It is worthy of
further study that whether there exists such an analytic solution or there
is something special in its framework if such a solution exists.

\bigskip 

\bigskip \appendix\textbf{Appendix A} \bigskip

Now we show briefly prove (\ref{69}). Set $\left\vert \phi \right\rangle
=\left\vert 0\right\rangle ,$%
\begin{eqnarray}
f_{st}(u,Av) &=&\underset{r=0}{\sum^{A-1}}\langle k,q+\frac{l\left(
s+r\right) }{A}|0\rangle \langle 0|k,q+\frac{lr}{A}\rangle \times e^{\frac{%
2\pi itr}{A}}\times e^{\frac{2\pi itq}{l}}  \notag \\
&=&\frac{1}{l\sqrt{\pi }}\underset{r}{\sum }\theta \left( \tilde{v}_{0}+%
\frac{\tilde{s}_{0}+r}{A},\frac{\tau }{A}\right) \theta \left( \tilde{v}_{0}+%
\frac{r}{A},-\frac{\tau ^{\ast }}{A}\right)  \notag \\
&& \times e^{\frac{2\pi itr}{A}}\times e^{\pi i\frac{\tau -\tau ^{\ast }}{A}%
u^{2}+2\pi i\frac{s}{A}u+2\pi itv}  \label{104}
\end{eqnarray}

\bigskip Note $\tilde{v}_{0}\neq v$, \ \ $\tilde{s}_{0}\neq s$ due to (\ref%
{80})(\ref{81}) and $-\frac{\tau ^{\ast }}{A}=\frac{\tau -1}{A}.$ In terms
of the definition of theta function, we expand the theta functions involved
in $F_{st}^{^{\prime }}$ defined as 
\begin{equation}
F_{st}^{^{\prime }}=\underset{r}{\sum }\theta \left( \tilde{v}_{0}+\frac{%
\tilde{s}_{0}+r}{A},\frac{\tau }{A}\right) \theta \left( \tilde{v}_{0}+\frac{%
r}{A},\frac{\tau -1}{A}\right) e^{\frac{2\pi itr}{A}}  \label{105}
\end{equation}%
in the Laurent series obtaining%
\begin{eqnarray*}
F_{st}^{^{\prime }} &=&\underset{r}{\sum }\left( \underset{m}{\sum }e^{\pi i%
\frac{\tau }{A}m^{2}}e^{2\pi im\left( \tilde{v}_{0}+\frac{\tilde{s}_{0}+r}{A}%
\right) }\right) \\
&&\times \left( \underset{m^{^{\prime }}}{\sum }e^{\pi i\frac{\tau -1}{A}%
m^{^{\prime }2}}e^{2\pi im^{^{\prime }}\left( \tilde{v}_{0}+\frac{r}{A}%
\right) }\right) \times e^{\frac{2\pi itr}{A}}.
\end{eqnarray*}%
After replacing variable $m^{^{\prime }}$ by $n-m$ $,$ we get $%
F^{\prime}_{st}(u,Av)$ as follow:

\begin{equation*}
F_{st}^{^{\prime }}=\underset{m,n}{\sum }e^{\frac{\pi i}{A}\{m^{2}\left(
2\tau -1\right) +n^{2}\left( \tau -1\right) -2mn\left( \tau -1\right)
\}}\times e^{2\pi in\tilde{v}_{0}}\times e^{2\pi i\frac{\tilde{s}_{0}}{A}%
m}\times \sum_{r=0}^{A-1}e^{2\pi i\frac{t+n}{A}r}.
\end{equation*}%
\ Due to 
\begin{equation*}
\sum_{r=0}^{A-1}e^{2\pi i\frac{t+n}{A}r}=%
\begin{cases}
A & \text{when $n=LA-t$~~~}L\in \text{{}}%
\mathbb{Z}
\text{,} \\ 
0 & \text{otherwise}.%
\end{cases}%
\end{equation*}%
and substituting $LA-t$ for $n$, (here $L$ runs over all integers$,$ after
some computation and arrangement), we have%
\begin{eqnarray*}
F_{st}^{^{\prime }} &=&A\sum_{m,L}e^{\frac{\pi i}{A}\left( 2\tau -1\right)
\left\{ \left( m-\frac{1}{2}\left( LA-t\right) \right) ^{2}+\frac{1}{4}%
\left( LA-t\right) ^{2}\right\} } \\
&&\times e^{-\frac{\pi i}{2A}\left\{ \left( LA\right) ^{2}-2LAt\right\}
}\times e^{-\frac{\pi i}{2A}t^{2}}\times e^{-\frac{\pi i}{A}\left( m-\frac{1%
}{2}\left( LA-t\right) \right) \left( t-2\tilde{s}_{0}\right) } \\
&&\times e^{-\frac{\pi i}{A}\left( \frac{1}{2}\left( LA-t\right) \right)
\left( t-2\tilde{s}_{0}\right) }\times e^{2\pi i\left( LA-t\right) \tilde{v}%
_{0}}\times e^{\pi imL}
\end{eqnarray*}%
Next we set $L=2h+\delta ,$ here\ $h\in 
\mathbb{Z}
,\delta =0,1$ and note the fact 
\begin{equation*}
e^{\pi imL}=e^{\pi im\left( 2h+\delta \right) }=e^{\pi im\delta }.
\end{equation*}%
We obtain the form of sum over three variables $N,m,\delta $ as follows,%
\begin{eqnarray}
F_{st}^{\prime } &=&A\sum_{\delta =0,1}\underset{m,h}{\sum }e^{\pi i\left( 
\frac{2\tau -1}{A}\right) \left\{ \left( m-\frac{1}{2}\left( \left(
2h+\delta \right) A-t\right) \right) ^{2}+\frac{1}{4}\left( \left( 2h+\delta
\right) A-t\right) ^{2}\right\} } \\
&&\times e^{-\frac{\pi i}{2A}\left\{ \left( 2h+\delta \right)
^{2}A^{2}-2\left( 2h+\delta \right) At\right\} }\times e^{-\frac{\pi i}{2A}%
t^{2}}\times e^{\pi im\delta }\times e^{2\pi i[(2h+\delta )A-t]\tilde{v}_{0}}
\notag \\
&&\times e^{-\frac{\pi i}{A}\left( m-\frac{1}{2}\left( \left( 2h+\delta
\right) A-t\right) \right) \left( t-2\tilde{s}_{0}\right) }\times e^{-\frac{%
\pi i}{2A}\left( \left( 2h+\delta \right) A-t\right) \left( t-2\tilde{s}%
_{0}\right) }
\end{eqnarray}%
After arrangement, we find the sum over $m$ and $n$ can be separated into
product of two theta functions. (note $\delta ^{2}=\delta $) 
\begin{eqnarray}
F_{st}^{^{\prime }} &=&A\underset{\delta }{\sum }\left( \underset{h}{\sum }%
e^{\pi iA\left( 2\tau -1\right) h^{2}}\times e^{2\pi ih\left[ \frac{\delta
A-t+2\tilde{s}_{0}}{2}+2A\tilde{v}_{0}+\left( 2\tau -1\right) \frac{\delta
A-t}{2}\right] }\right) \\
&&\times \left( \underset{m}{\sum }e^{\pi i\frac{2\tau -1}{A}\left(
m-hA\right) ^{2}}\times e^{2\pi i\left( m-hA\right) \left[ \frac{\delta A-t+2%
\tilde{s}_{0}}{2A}-\left( \frac{2\tau -1}{A}\right) \left( \frac{\delta A-t}{%
2}\right) \right] }\right) \\
&&\times e^{2\pi i\frac{2\tau -1}{A}\left( \frac{\delta A-t}{2}\right)
^{2}}\times e^{-\frac{\pi i}{2A}\left( \delta A-t\right) ^{2}}\times e^{2\pi
i\left( \delta A-t\right) \tilde{v}_{0}} \\
&=&\underset{\delta =0,1}{\sum }A\theta \left( \frac{\left( \delta
A-t\right) }{A}\left( 1-\tau \right) +\frac{\tilde{s}_{0}}{A},\frac{2\tau -1%
}{A}\right) \\
&&\times \theta \left( \left( \delta A-t\right) \tau +\tilde{s}_{0}+2A\tilde{%
v}_{0},A\left( 2\tau -1\right) \right) \times e^{2\pi i\frac{2\tau -2}{A}%
\left( \frac{\delta A-t}{2}\right) ^{2}}\times e^{2\pi i\left( \delta
A-t\right) \tilde{v}_{0}} \\
&\equiv &\underset{\delta }{\sum }A\theta \left( z,\frac{2\tau -1}{A}\right)
\theta \left( w,A\left( 2\tau -1\right) \right) e^{2\pi i\phi },
\end{eqnarray}%
where 
\begin{eqnarray}
\phi &=&\frac{\tau -1}{2A}\left( \delta A-t\right) ^{2}+\left( \delta
A-t\right) \tilde{v}_{0}, \\
z &=&\left( \frac{\delta A-t}{A}\right) \left( 1-\tau \right) +\frac{\tilde{s%
}_{0}}{A}, \\
w &=&\left( \delta A-t\right) \tau +\tilde{s}_{0}+2A\tilde{v}_{0}.
\end{eqnarray}%
Having known that 
\begin{equation*}
f_{st}=F_{st}^{^{\prime }}e^{\pi i\frac{\tau -\tau ^{\ast }}{A}u^{2}+2\pi i%
\frac{s}{A}u+2\pi itv},
\end{equation*}%
from (\ref{104})(\ref{105}) and $\tilde{v}_{0}=v+\frac{\tau ^{\ast }u}{A}$
and $\frac{\tilde{s}_{0}}{A}=\frac{s}{A}+\frac{(2\tau -1)u}{A}~,~$we obtain 
\begin{equation*}
f_{st}=\underset{\delta }{\sum }A\theta \left( z^{\prime },\frac{2\tau -1}{A}%
\right) \theta \left( w,A\left( 2\tau -1\right) \right) e^{2\pi i\phi
^{^{\prime }}},
\end{equation*}%
where%
\begin{eqnarray*}
z^{\prime } &=&z-t\frac{2\tau -1}{A}=-\delta \tau -\frac{\tau }{A}t+\frac{s}{%
A}+\frac{2\tau -1}{A}u, \\
w &=&(\delta A-t)\tau +s+2Av+u, \\
\phi ^{^{\prime }} &=&\delta A\frac{\tau -1}{2}+\delta (Av-(\tau -1)u)+\frac{%
2\tau -1}{2A}u^{2}+\frac{s}{A}u, \\
&&-\frac{\tau }{A}tu-\frac{st}{A}+\frac{\tau }{2A}t^{2}.
\end{eqnarray*}%
This is the formula (\ref{69}).

\bigskip

\appendix\textbf{Appendix B} \bigskip

Now we would like to derive the general form of the projectors. We have the
inner product of $<k,q|$ and the coherent state $|z>$ in paper \cite{Deng}%
\begin{equation*}
<k,q|\left( z^{\prime }\right) >=\frac{1}{\sqrt{l}\pi ^{1/4}}\theta (\frac{q+%
\frac{\tau }{\tau _{2}}k-i\sqrt{2}z^{\prime }}{l},\frac{\tau }{A})e^{-\frac{%
\tau }{2i\tau _{2}}k^{2}+ikq+\sqrt{2}kz^{\prime }-(z^{\prime 2}+z^{\prime }%
\bar{z}^{\prime })/2},
\end{equation*}%
where $a|\left( z^{\prime }\right) >=z^{\prime }|\left( z^{\prime }\right) >.
$ Let $\frac{l}{i\sqrt{2}}z=z^{\prime },|\left( z^{\prime }\right) >=|z>,$
we have $a|z>=\frac{l}{i\sqrt{2}}z|z>.$(In \cite{Deng}, the coherent state
is denoted by $|z>$ with $a|z>=z|z>$ which is the same as $|\left( z\right) >
$ in this paper, we have given another implication to $|z>$ in this paper).
Substitute the above formula into $G_{ss^{\prime }}(u,v)$ 
\begin{eqnarray*}
&&G_{ss^{\prime }}(u,v,z_{1},z_{2}^{\ast }) \\
&=&<k,q+\frac{ls}{A}|z_{1}><z_{2}|k,q+\frac{ls^{\prime }}{A}> \\
&=&\frac{1}{l\sqrt{\pi }}\theta (v+\frac{\tau u}{A}+\frac{s}{A}-z_{1},\frac{%
\tau }{A})\theta (v+\frac{\tau ^{\ast }u}{A}+\frac{s^{\prime }}{A}%
-z_{2}^{\ast },\frac{-\tau ^{\ast }}{A}) \\
&&\times e^{\pi i\frac{\tau -\tau ^{\ast }}{A}u^{2}+2\pi iu(\frac{%
s-s^{\prime }}{A}-z_{1}+z_{2}^{\ast })}\times e^{\frac{l^{2}}{4}%
(z_{1}^{2}+z_{2}^{\ast 2}+z_{1}z_{1}^{\ast }+z_{2}z_{2}^{\ast })}.
\end{eqnarray*}%
\begin{eqnarray}
f_{st}(u,Av,z_{1},z_{2}^{\ast }) &\equiv
&\sum_{r=0}^{A-1}G_{s+r,r}(u,v,z_{1},z_{2}^{\ast })\times e^{2\pi it(\frac{r%
}{A}+v)}  \notag \\
&=&\sum_{r}\frac{1}{l\sqrt{\pi }}\theta (v+\frac{\tau u}{A}+\frac{s+r}{A}%
-z_{1},\frac{\tau }{A})\theta (v+\frac{r}{A}+\frac{u\tau ^{\ast }}{A}%
-z_{2}^{\ast },\frac{-\tau ^{\ast }}{A})\times e^{2\pi itr/A}  \notag \\
&&\times e^{\pi i\frac{\tau -\tau ^{\ast }}{A}u^{2}+2\pi iu(\frac{s}{A}%
-z_{1}+z_{2}^{\ast })+2\pi itv}\times e^{\frac{l^{2}}{4}(z_{1}^{2}+z_{2}^{%
\ast 2}+z_{1}z_{1}^{\ast }+z_{2}z_{2}^{\ast })}  \notag
\end{eqnarray}%
Define $\tilde{v},\tilde{s}$ as%
\begin{equation}
\tilde{v}+\frac{\tilde{s}}{A}=v+\frac{\tau u}{A}+\frac{s}{A}-z_{1},~~\tilde{v%
}=v+\frac{\tau ^{\ast }u}{A}+z_{2}^{\ast }  \label{65}
\end{equation}%
From (\ref{65}) we have for $\tau =e^{\frac{\pi i}{3}}$ 
\begin{eqnarray}
u &=&\frac{-i}{\sqrt{3}}(\tilde{s}-s+A(z_{1}-z_{2}^{\ast }))  \label{66} \\
Av &=&A\tilde{v}+\frac{\tilde{s}-s}{2}(1+\frac{i}{\sqrt{3}})+\frac{Az_{1}}{2}%
(1+\frac{i}{\sqrt{3}})+\frac{Az_{2}^{\ast }}{2}(1-\frac{i}{\sqrt{3}}).
\label{67}
\end{eqnarray}%
In terms of new variable $\tilde{v}$ and $\tilde{s}$ ,we have%
\begin{eqnarray}
f_{st}(u,Av,z_{1},z_{2}^{\ast }) &=&\frac{A}{l\sqrt{\pi }}\sum_{\delta
}\theta ((\delta A-t)(1-\tau )+\frac{\tilde{s}}{A},\frac{2\tau -1}{A}) 
\notag \\
&&\times \theta ((\delta A-t)\tau +2A\tilde{v}+\tilde{s},A(2\tau -1)\times
e^{2\pi i\frac{\tau -1}{2A}\left( \delta A-t\right) ^{2}}  \notag \\
&&\times e^{2\pi i\delta A\tilde{v}}\times e^{2\pi it\left( \frac{\tilde{s}-s%
}{2A}(1+\frac{i}{\sqrt{3}})+\frac{z_{1}}{2}(1+\frac{i}{\sqrt{3}})+\frac{%
z_{2}^{\ast }}{2}(1-\frac{i}{\sqrt{3}})\right) }  \notag \\
&&\times e^{\frac{\pi }{\sqrt{3}A}(\tilde{s}^{2}-s^{2}+2sA(z_{1}-z_{2}^{\ast
}))}\times e^{\frac{l^{2}}{4}(2z_{1}z_{2}^{\ast }+z_{1}z_{1}^{\ast
}+z_{2}z_{2}^{\ast })}.  \label{64}
\end{eqnarray}%
Taking $z_{1}=z_{2}=0$ in Eq.(\ref{64}) and denote $\tilde{v}_{0}=\tilde{v}%
(z_{1}=z_{2}=0),\tilde{s}_{0}=\tilde{s}(z_{1}=z_{2}=0)$ we get 
\begin{eqnarray}
f_{st}(u,Av,z_{1},z_{2}^{\ast })_{z_{1}=z_{2}=0} &=&\frac{A}{l\sqrt{\pi }}%
\sum_{\delta }\theta \left( (\delta A-t)(1-\tau )+\frac{\tilde{s}_{0}}{A},%
\frac{2\tau -1}{A}\right)   \notag \\
&&\theta \left( (\delta A-t)\tau +2A\tilde{v}_{0}+\tilde{s}_{0},A(2\tau
-1\right)   \notag \\
&&\times e^{2\pi i\frac{\tau -1}{2A}\left( \delta A-t\right) ^{2}}\times
e^{2\pi i\delta A\tilde{v}_{0}}\times e^{2\pi it\frac{\tilde{s}_{0}-s}{2A}(1+%
\frac{i}{\sqrt{3}})}\times e^{\frac{\pi }{\sqrt{3}A}(\tilde{s}%
_{0}^{2}-s^{2})}.  \notag \\
&&  \label{112}
\end{eqnarray}%
On the other hand, when $z_{1}=z_{2}=0,$ $f_{st}(u,Av,0,0)=f_{st}(u,Av).$ We
have from (\ref{80})(\ref{94})(\ref{95}) 
\begin{eqnarray}
f_{st}(u,Av) &=&c_{0}\theta \left( A\tilde{v}_{0}+\frac{\tilde{s}_{0}-s}{2}%
(1+\frac{i}{\sqrt{3}})+\alpha \right)   \notag \\
&&\times \theta \left( A\tilde{v}_{0}+\frac{\tilde{s}_{0}-s}{2}(1-\frac{i}{%
\sqrt{3}})+\beta \right) \theta \left( \frac{-i}{\sqrt{3}}(\tilde{s}%
_{0}-s)+\gamma \right)   \notag \\
&&+c_{1}\theta \left( A\tilde{v}_{0}+\frac{\tilde{s}_{0}-s}{2}(1+\frac{i}{%
\sqrt{3}})+\alpha +\frac{1}{2}\right)   \notag \\
&&\times \theta \left( A\tilde{v}_{0}+\frac{\tilde{s}_{0}-s}{2}(1-\frac{i}{%
\sqrt{3}})+\beta -\frac{1}{2}\right) \theta \left( \frac{-i}{\sqrt{3}}(%
\tilde{s}_{0}-s)+\gamma +\frac{1}{2}\right) ,  \label{113}
\end{eqnarray}%
where we have from (\ref{103}) (\ref{106})%
\begin{eqnarray}
c_{0} &=&\frac{A}{l\sqrt{\pi }}e^{2\pi i(\phi _{st}+\phi _{\delta
=0})}\theta \left( \delta \frac{\left( 1-A\right) }{2}\frac{2\tau -1}{A},%
\frac{2\tau -1}{A}\right)   \notag  \label{102} \\
&&\times \theta \left( (\frac{1}{2}-\frac{m}{3})A(2\tau -1)+\frac{1}{2}%
,A(2\tau -1)\right)   \notag \\
&&\div \left\{ \theta \left( \frac{A\tau +1}{2}+\frac{1}{2}\right) \theta
\left( \frac{A\tau +1}{2}-\frac{m}{3}A(2\tau -1),A\tau \right) \theta \left( 
\frac{1-A}{2}-\frac{m}{3}A(2\tau -1)\right) \right\} ,  \notag \\
&&
\end{eqnarray}%
\begin{eqnarray}
c_{1} &=&\frac{A}{l\sqrt{\pi }}e^{2\pi i(\phi _{st}+\phi _{\delta
=1})}\theta \left( \delta \frac{\left( 1-A\right) }{2}\frac{2\tau -1}{A},%
\frac{2\tau -1}{A}\right)   \notag  \label{101} \\
&&\times \theta \left( (\frac{1}{2}-\frac{m}{3})A(2\tau -1)+\frac{1}{2}%
,A(2\tau -1)\right)   \notag \\
&&\div \left\{ \theta \left( \frac{A\tau +1}{2}+\frac{1}{2}\right) \theta
\left( \frac{A\tau +1}{2}-\frac{m}{3}A(2\tau -1),A\tau \right) \theta \left( 
\frac{1-A}{2}-\frac{m}{3}A(2\tau -1)+\frac{1}{2}\right) \right\} .  \notag \\
&&
\end{eqnarray}%
Thus as an analytic function of $A\tilde{v}_{0}$ and $\tilde{s}_{0}$, the
right-handed side of (\ref{112}) equals to the right-handed side of (\ref%
{113}) holds for all $u$ and $v$ (see section 3) and it is an identity.
Through the transformation of $A\tilde{v}_{0}\rightarrow A\tilde{v},~~\tilde{%
s}_{0}\rightarrow \tilde{s}$, we get 
\begin{eqnarray}
&&\frac{A}{l\sqrt{\pi }}\sum_{\delta }\theta \left( (\delta A-t)(1-\tau )+%
\frac{\tilde{s}}{A},\frac{2\tau -1}{A}\right)   \notag \\
&&\theta \left( (\delta A-t)\tau +2A\tilde{v}+\tilde{s},A(2\tau -1\right)  
\notag \\
&&\times e^{2\pi i\frac{\tau -1}{2A}\left( \delta A-t\right) ^{2}}\times
e^{2\pi i\delta A\tilde{v}}\times e^{2\pi it\frac{\tilde{s}-s}{2A}(1+\frac{i%
}{\sqrt{3}})}\times e^{\frac{\pi }{\sqrt{3}A}(\tilde{s}^{2}-s^{2})}  \notag
\\
&=&c_{0}\theta \left( A\tilde{v}+\frac{\tilde{s}-s}{2}(1+\frac{i}{\sqrt{3}}%
)+\alpha \right) \theta \left( A\tilde{v}+\frac{\tilde{s}-s}{2}(1-\frac{i}{%
\sqrt{3}})+\beta \right) \theta \left( -\frac{i}{\sqrt{3}}\frac{\tilde{s}-s}{%
2}+\gamma \right)   \notag \\
&&+c_{1}\theta \left( A\tilde{v}+\frac{\tilde{s}-s}{2}(1+\frac{i}{\sqrt{3}}%
)+\alpha +\frac{1}{2}\right) \theta \left( A\tilde{v}+\frac{\tilde{s}-s}{2}%
(1-\frac{i}{\sqrt{3}})+\beta -\frac{1}{2}\right)   \notag \\
&&\times \theta \left( -\frac{i}{\sqrt{3}}\frac{\tilde{s}-s}{2}+\gamma +%
\frac{1}{2}\right) \equiv f_{st}^{\prime }.  \label{68}
\end{eqnarray}%
Observing the right-hand side of (\ref{64}) compared with the left-hand side
of (\ref{68}), it is easy to find%
\begin{eqnarray*}
f_{st}(u,Av,z_{1},z_{2}^{\ast }) &=&f_{st}^{\prime }\times e^{2\pi it[(z_{1}(%
\frac{1}{2}+\frac{\sqrt{3}i}{6})+z_{2}^{\ast }(\frac{1}{2}-\frac{\sqrt{3}i}{6%
})]}\times e^{2\pi is(\frac{-\sqrt{3}i}{3}z_{1}+\frac{\sqrt{3}i}{3}%
z_{2}^{\ast })} \\
&&\times e^{\frac{l^{2}}{4}(2z_{1}z_{2}^{\ast }+z_{1}z_{1}^{\ast
}+z_{2}z_{2}^{\ast })}
\end{eqnarray*}%
Substituting (\ref{65}) into (\ref{68}), we get%
\begin{eqnarray}
f_{st}(u,Av,z_{1},z_{2}^{\ast }) &=&c_{0}\theta \left( Av-Az_{1}(\frac{1}{2}+%
\frac{\sqrt{3}i}{6})-Az_{2}^{\ast }(\frac{1}{2}-\frac{\sqrt{3}i}{6})+\alpha
\right)   \notag \\
&&\times \theta \left( Av+u-Az_{1}(\frac{1}{2}-\frac{\sqrt{3}i}{6}%
)-Az_{2}^{\ast }(\frac{1}{2}+\frac{\sqrt{3}i}{6})+\beta \right)   \notag \\
&&\times \theta \left( u-Az_{1}(-\frac{i}{\sqrt{3}})-Az_{2}^{\ast }(\frac{i}{%
\sqrt{3}})+\gamma \right)   \notag \\
&&+c_{1}\theta \left( Av-Az_{1}(\frac{i}{\sqrt{3}})-Az_{2}^{\ast }(\frac{1}{2%
}-\frac{\sqrt{3}i}{6})+\alpha +\frac{1}{2}\right)   \notag \\
&&\times \theta \left( Av+u-Az_{1}(\frac{1}{2}-\frac{\sqrt{3}i}{6}%
)-Az_{2}^{\ast }(\frac{i}{\sqrt{3}})+\beta -\frac{1}{2}\right)   \notag \\
&&\times \theta \left( u-Az_{1}(-\frac{i}{\sqrt{3}})-Az_{2}^{\ast }(\frac{i}{%
\sqrt{3}})+\gamma +\frac{1}{2}\right)   \notag \\
&&\times e^{2\pi it[(z_{1}(\frac{1}{2}+\frac{\sqrt{3}i}{6})+z_{2}^{\ast }(%
\frac{1}{2}-\frac{\sqrt{3}i}{6})]}\times e^{2\pi is(\frac{-\sqrt{3}i}{3}%
z_{1}+\frac{\sqrt{3}i}{3}z_{2}^{\ast })}\times e^{\frac{l^{2}}{4}%
(2z_{1}z_{2}^{\ast }+z_{1}z_{1}^{\ast }+z_{2}z_{2}^{\ast })}  \notag \\
&&.
\end{eqnarray}%
Take the transformation 
\begin{equation*}
u\longrightarrow -Av,~~Av\longrightarrow -\frac{A}{2}+u+Av.
\end{equation*}%
The $f_{st}$ is transformed into $\bar{f}_{st}$ 
\begin{eqnarray}
\bar{f}_{st}(u,Av,z_{1},z_{2}^{\ast }) &=&c_{0}(s,t)\theta \left( Av-Az_{1}(%
\frac{i}{\sqrt{3}})-Az_{2}^{\ast }(-\frac{i}{\sqrt{3}})+\alpha
_{t,t-s}\right)   \notag \\
&&\times \theta \left( Av+u-Az_{1}(\frac{1}{2}+\frac{\sqrt{3}i}{6}%
)-Az_{2}^{\ast }(\frac{1}{2}-\frac{\sqrt{3}i}{6})+\beta _{t,t-s}\right)  
\notag \\
&&\times \theta \left( u-Az_{1}(\frac{1}{2}-\frac{\sqrt{3}i}{6}%
)-Az_{2}^{\ast }(\frac{1}{2}+\frac{\sqrt{3}i}{6})+\gamma _{t,t-s}\right)  
\notag \\
&&+c_{1}(s,t)\theta \left( Av-Az_{1}(\frac{i}{\sqrt{3}})-Az_{2}^{\ast }(-%
\frac{i}{\sqrt{3}})+\alpha _{t,t-s}+\frac{1}{2}\right)   \notag \\
&&\times \theta \left( Av+u-Az_{1}(\frac{1}{2}+\frac{\sqrt{3}i}{6}%
)-Az_{2}^{\ast }(\frac{1}{2}-\frac{\sqrt{3}i}{6})+\beta _{t,t-s}-\frac{1}{2}%
\right)   \notag \\
&&\times \theta \left( u-Az_{1}(\frac{1}{2}-\frac{\sqrt{3}i}{6}%
)-Az_{2}^{\ast }(\frac{1}{2}+\frac{\sqrt{3}i}{6})+\gamma _{t,t-s}+\frac{1}{2}%
\right)   \notag \\
&&\times e^{2\pi i(t-s)(\frac{\sqrt{3}i}{3}z_{1}-\frac{\sqrt{3}i}{3}%
z_{2}^{\ast })}\times e^{2\pi it[(z_{1}(\frac{1}{2}-\frac{\sqrt{3}i}{6}%
)+z_{2}^{\ast }(\frac{1}{2}+\frac{\sqrt{3}i}{6})]}  \notag \\
&&\times e^{\frac{l^{2}}{4}(2z_{1}z_{2}^{\ast }+z_{1}z_{1}^{\ast
}+z_{2}z_{2}^{\ast })}.
\end{eqnarray}%
Again we let $s^{\prime }=t,t^{\prime }=t-s$ and change $z_{1}$ and $z_{2}$
into $z_{1}^{\prime }=e^{\frac{2\pi i}{6}}z_{1}$ and $z_{2}^{\prime }=e^{-%
\frac{2\pi i}{6}}z_{2}$ and rewrite $c_{j}(st)$ as $e^{2\pi i\phi
_{st}}c_{j}(00).$(see (\ref{102}) and (\ref{101})and the definition of $\phi
_{st},\phi _{\delta }$ in (\ref{55})) We have

\begin{eqnarray}
\bar{f}_{st}(u,Av,z_{1},z_{2}^{\ast }) &=&e^{2\pi i\phi
_{st}}c_{0}(0,0)\theta \left( Av-Az_{1}^{\prime }(\frac{1}{2}+\frac{\sqrt{3}i%
}{6})-Az_{2}^{\prime \ast }(\frac{1}{2}-\frac{\sqrt{3}i}{6})+\alpha
_{s^{\prime },t^{\prime }}\right)  \notag  \label{83} \\
&&\times \theta \left( Av+u-Az_{1}^{\prime }(\frac{1}{2}-\frac{\sqrt{3}i}{6}%
)-Az_{2}^{\prime \ast }(\frac{1}{2}+\frac{\sqrt{3}i}{6})+\beta _{s^{\prime
},t^{\prime }}\right)  \notag \\
&&\times \theta \left( u-Az_{1}^{\prime }(-\frac{i}{\sqrt{3}}%
)-Az_{2}^{\prime \ast }(\frac{i}{\sqrt{3}})+\gamma _{s^{\prime },t^{\prime
}}\right)  \notag \\
&&+e^{2\pi i\phi _{st}}c_{1}(0,0)\theta \left( Av-Az_{1}^{\prime }(\frac{1}{2%
}+\frac{\sqrt{3}i}{6})-Az_{2}^{\prime \ast }(\frac{1}{2}-\frac{\sqrt{3}i}{6}%
)+\alpha _{s^{\prime },t^{\prime }})+\frac{1}{2}\right)  \notag \\
&&\times \theta \left( Av+u-Az_{1}^{\prime }(\frac{1}{2}-\frac{\sqrt{3}i}{6}%
)-Az_{2}^{\prime \ast }(\frac{1}{2}+\frac{\sqrt{3}i}{6})+\beta _{s^{\prime
},t^{\prime }}-\frac{1}{2}\right)  \notag \\
&&\times \theta \left( u-Az_{1}^{\prime }(-\frac{i}{\sqrt{3}}%
)-Az_{2}^{\prime \ast }(\frac{i}{\sqrt{3}})+\gamma _{s^{\prime },t^{\prime
}}+\frac{1}{2}\right)  \notag \\
&&\times e^{2\pi it^{\prime }[(z_{1}^{\prime }(\frac{1}{2}+\frac{\sqrt{3}i}{6%
})+z_{2}^{\prime \ast }(\frac{1}{2}-\frac{\sqrt{3}i}{6})]}\times e^{\frac{%
l^{2}}{4}(2z_{1}^{\prime }z_{2}^{\prime \ast }+z_{1}^{\prime }z_{1}^{\prime
\ast }+z_{2}^{\prime }z_{2}^{\prime \ast })}\times e^{2\pi is^{\prime }(%
\frac{-\sqrt{3}i}{3}z_{1}^{\prime }+\frac{\sqrt{3}i}{3}z_{2}^{\prime \ast
})}.  \notag \\
&&
\end{eqnarray}%
It is easy to check that 
\begin{equation*}
e^{2\pi i\phi _{st}}=c^{2st-t^{2}}e^{2\pi i\phi _{s^{\prime },t^{\prime }}},
\end{equation*}%
where%
\begin{equation*}
\phi _{st}=\frac{\sqrt{3}i}{6A}\left( s^{2}+t^{2}-st\right) -\frac{1}{2A}st.
\end{equation*}%
Therefore, we have%
\begin{equation}
f_{s^{\prime }t^{\prime }}(u,Av,z_{1}^{\prime },z_{2}^{\prime \ast
})=c^{-2st+t^{2}}\bar{f}_{st}(u,Av,z_{1},z_{2}^{\ast }).  \label{114}
\end{equation}%
Finally we check the $\psi _{s,t}(u,Av)$ constructed from $f_{st}$ for the
covariant condition (\ref{44}),%
\begin{equation}
c^{-2st+t^{2}}\psi _{st}(-Av,-\frac{A}{2}+u+Av)=\psi _{t,t-s}(u,Av)=\psi
_{s^{\prime },t^{\prime }}(u,Av).  \label{85}
\end{equation}%
Since%
\begin{eqnarray*}
\psi _{s^{\prime },t^{\prime }}(u,Av) &=&\frac{f_{s^{\prime }t^{\prime
}}(u,v)}{Af_{00}(u,v)} \\
&=&\frac{\int d^{2}z_{1}^{\prime }d^{2}z_{2}^{\prime }f_{s^{\prime
}t^{\prime }}(u,Av,z_{1}^{\prime },z_{2}^{\prime \ast })F_{1}(z_{1}^{\prime
})F_{2}(z_{2}^{\prime })^{\ast }}{A\int d^{2}z_{1}^{\prime
}d^{2}z_{2}^{\prime }f_{00}(u,Av,z_{1}^{\prime },z_{2}^{\prime \ast
})F_{1}(z_{1}^{\prime })F_{2}(z_{2}^{\prime })^{\ast }} \\
&=&\frac{c^{-2st+t^{2}}\int d^{2}z_{1}^{\prime }d^{2}z_{2}^{\prime }\bar{f}%
_{st}(u,Av,z_{1},z_{2}^{\ast })F_{1}(z_{1}^{\prime })F_{2}(z_{2}^{\prime
})^{\ast }}{A\int d^{2}z_{1}^{\prime }d^{2}z_{2}^{\prime }\bar{f}%
_{00}(u,Av,z_{1},z_{2}^{\ast })F_{1}(z_{1}^{\prime })F_{2}(z_{2}^{\prime
})^{\ast }}
\end{eqnarray*}%
Based on (\ref{25.4}) and (\ref{111}) 
\begin{eqnarray*}
&&\psi _{st}(-Av,-\frac{A}{2}+u+Av) \\
&=&\frac{\int d^{2}z_{1}d^{2}z_{2}\bar{f}_{st}(u,Av,z_{1},z_{2}^{\ast
})F_{1}(z_{1})F_{2}(z_{2}^{\ast })}{A\int d^{2}z_{1}d^{2}z_{2}\bar{f}%
_{00}(u,Av,z_{1},z_{2}^{\ast })F_{1}(z_{1})F_{2}(z_{2}^{\ast })}.
\end{eqnarray*}%
It is obvious that the equation (\ref{85}) holds.

\end{document}